  \documentclass[structabstract]{aa}
\usepackage{graphicx}
\usepackage{txfonts}
\usepackage{natbib}
\usepackage{ulem}

\bibpunct[]{(}{)}{;}{a}{}{,}

\begin{document}
   \title{Laboratory spectroscopic study of isotopic thioformaldehyde, H$_2$CS, 
          and determination of its equilibrium structure\thanks{Transition 
          frequencies from this work as well as related data from earlier work 
          are given for each isotopic species as supplementary material. 
          Given are also quantum numbers, uncertainties, and residuals between 
          measured frequencies and those calculated from the final sets of 
          spectroscopic parameters. The data are available at CDS via anonymous 
          ftp to cdsarc.u-strasbg.fr (130.79.128.5) or via 
          http://cdsweb.u-strasbg.fr/cgi-bin/qcat?J/A+A/621/A143}}

   \author{Holger S.~P. M{\"u}ller\inst{1}
           \and
           Atsuko Maeda\inst{2}
           \and
           Sven Thorwirth\inst{1}
           \and
           Frank Lewen\inst{1}
           \and
           Stephan Schlemmer\inst{1}
           \and
           Ivan R. Medvedev\inst{2,\thanks{Current address: Department of Physics, Wright State University, Dayton, OH 45435 USA.}}
           \and
           Manfred Winnewisser\inst{2}
           \and
           Frank C. De Lucia\inst{2}
           \and
           Eric Herbst\inst{2,\thanks{Current address: Departments of Chemistry and Astronomy, University of Virginia, Charlottesville, VA 22904, USA.}}
           }

   \institute{I.~Physikalisches Institut, Universit{\"a}t zu K{\"o}ln,
              Z{\"u}lpicher Str. 77, 50937 K{\"o}ln, Germany\\
              \email{hspm@ph1.uni-koeln.de}
              \and
              Department of Physics, The Ohio State University, Columbus, 
              OH~43210-1107, USA
              }

   \date{Received 26 October 2018 / Accepted 22 November 2018}
 
  \abstract
{Thioformaldehyde is an abundant molecule in various regions of the interstellar medium. 
However, available laboratory data limit the accuracies of calculated transition frequencies 
in the submillimeter region, in particular for minor isotopic species.}
{We aim to determine spectroscopic parameters of isotopologs of H$_2$CS that are accurate 
enough for predictions well into the submillimeter region.}
{We investigated the laboratory rotational spectra of numerous isotopic species in natural 
isotopic composition almost continuously between 110 and 377~GHz. Individual lines were 
studied for most species in two frequency regions between 566 and 930~GHz. Further data 
were obtained for the three most abundant species in the 1290$-$1390~GHz region.}
{New or improved spectroscopic parameters were determined for seven isotopic species. 
Quantum-chemical calculations were carried out to evaluate the differences between 
ground state and equilibrium rotational parameters to derive semi-empirical equilibrium 
structural parameters.}
{The spectroscopic parameters are accurate enough for predictions well above 1~THz with the 
exception of H$_2^{13}$C$^{34}$S where the predictions should be reliable to around 700~GHz.}

\authorrunning{H.~S.~P. M{\"u}ller et al.}
\titlerunning{Laboratory spectroscopic study of isotopic H$_2$CS}

\maketitle
\hyphenation{For-schungs-ge-mein-schaft}

\section{Introduction}
\label{intro}

Thioformaldehyde, H$_2$CS, was among the molecules detected early in space, namely in the giant 
high-mass starforming region Sagittarius~B2 near the Galactic center \citep{H2CS_det_1973}. 
The molecule was also detected in dark clouds, such as TMC-1 and L134N \citep{H2CS_dark-clouds_1989}, 
and in the circumstellar envelope of the C-rich asymptotic giant branch (AGB) star CW~Leonis, 
also known as IRC +10216 \citep{H2CS_etc_CW-Leo_2008}. Concerning solar system objects, 
H$_2$CS was detected in the comet Hale Bopp \citep{H2CS_Hale-Bopp_1997}. Furthermore, 
it was detected in nearby galaxies, such as the Large Magellanic Cloud \citep{LMC_SMC_1999} 
and NGC253 \citep{NGC253_2005}, and also in more distant galaxies, such as the $z = 0.89$ 
foreground galaxy in the direction of the blazar PKS 1830$-$211 \citep{PKS1830_4mm_2011}. 
Several isotopic species were detected as well; H$_2$C$^{34}$S \citep{H2CS-34_1985}, 
H$_2 ^{13}$CS \citep{SgrB2_survey_1986}, HDCS \citep{HDCS_rot_det_1997}, and even D$_2$CS 
\citep{D2CS_det_2005}; unlabeled atoms refer to $^{12}$C and $^{32}$S.

Spectroscopic identifications of thioformaldehyde were based on molecular parameters which 
were obtained to a large extent from laboratory rotational spectroscopy. The first results 
were reported by \citet{H2CS_rot_1970} followed by additional measurements of H$_2$CS and, 
to a much lesser extent, of H$_2$C$^{34}$S, H$_2^{13}$CS, and D$_2$CS up to 70~GHz 
\citep{H2CS_isos_rot_1971}. \citet{H2CS_rot_1972} measured further transitions of H$_2$CS 
up to 244~GHz. \citet{H2CS_isos_1982} carried out microwave measurements of several minor 
thioformaldehyde isotopologs and determined dipole moments for H$_2$CS and D$_2$CS; 
a very accurate H$_2$CS dipole moment was reported by \citet{H2CS_dip_1977}. 
\citet{H2CS_HFS_1987} investigated the $^{33}$S and $^{13}$C hyperfine structure (HFS) from 
microwave transitions. \citet{HDCS_rot_det_1997} determined HDCS transition frequencies 
from the millimeter to the lower submillimeter regions. Additional, though less accurate 
data for H$_2$CS and H$_2$C$^{34}$S were obtained in a far-infrared study of thioformaldehyde 
\citep{H2CS_FIR_nu2_1993} and from the A$-$X electronic spectrum of H$_2$CS \citep{H2CS_A-X_1994}. 
Further, quite accurate transition frequencies of HDCS and D$_2$CS were obtained from 
radio-astronomical observations \citep{D2CS_det_2005}.

The need for higher frequency data was apparent in molecular line surveys of Orion~KL 
carried out with the Caltech Submillimeter Observatory (CSO) on Mauna Kea, Hawaii covering 
325$-$360~GHz \citep{Orion-KL-survey_1997} and 607$-$725~GHz \citep{Orion-KL-survey_2001}, 
and those carried out with the \textit{Odin} satellite covering 486$-$492~GHz and 
541$-$577~GHz \citep{Orion-KL-survey_2007-1,Orion-KL-survey_2007-2}, the one carried out 
with the \textit{Herschel} satellite covering 480$-$1280~GHz and 1426$-$1907~GHz 
\citep{Orion-KL_HIFI_2014}, and the the \textit{Herschel} molecular line survey 
of Sagittarius~B2(N) \citep{SgrB2N_HIFI_2014}. 
Moreover, the identification of H$_2$C$^{34}$S in the Protostellar Interferometric 
Line Survey (PILS) of IRAS 16293$-$2422 with the Atacama Large Millimeter/submillimeter 
Array (ALMA) between 329 and 363~GHz \citep{sulfur_IRAS16293_2018} may have been hampered 
by insufficient accuracies of some of the rotational transitions.

The apparent lack of accuracy in the H$_2$CS rest-frequencies in the higher frequency 
CSO survey \citep{Orion-KL-survey_2001} and a specific request from a member of the 
\textit{Odin} team to one of us (E.H.) initiated our study covering 110$-$377~GHz, 
566$-$670~GHz, and 848$-$930~GHz. An account on the H$_2$CS data in the ground vibrational 
state has been given by \citet{H2CS_rot_2008}. Later, we extended measurements to 
the 1290$-$1390~GHz region. Here, we report on the ground state rotational data of seven 
isotopic species, H$_2$CS, H$_2$C$^{33}$S, H$_2$C$^{34}$S, H$_2$C$^{36}$S, H$_2^{13}$CS, 
H$_2^{13}$C$^{34}$S, and HDCS obtained from samples in natural isotopic composition. 
The derived, often greatly improved, spectroscopic parameters permit predictions of 
accurate rest-frequencies well above 1~THz except for H$_2^{13}$C$^{34}$S, where the 
experimental data are more limited. The rotational parameters of these isotopologs 
plus a set of redetermined values for D$_2$CS combined with vibration-rotation parameters 
from quantum-chemical calculations were used to derive equilibrium structural parameters.

\section{Laboratory spectroscopic details}
\label{exptl}

We employed the Fast Scan Submillimeter-wave Spectroscopic Technique (FASSST) of The Ohio 
State University (OSU) to cover the 110$-$377~GHz range with a small gap at 190$-$200~GHz 
\citep{FASSST_1997,DEE_rot_FASSST_2004}. Additionally, we used two different spectrometer 
systems at the Universit{\"a}t zu K{\"o}ln to record higher frequency transitions up to 
almost 1.4~THz \citep{BWO-review_1994,BWO-review_1995,CH3SH_rot_THz-chain_2012}.

The FASSST system employs backward wave oscillators (BWOs) as sources; in the present 
study one that covered about 110$-$190~GHz and two spanning the region of 200$-$377~GHz. 
The frequency of each BWO was swept quickly so that a wide frequency range ($\sim$90~GHz) 
can be measured in a short period and any voltage instability of the BWOs can be overcome. 
The frequency of the FASSST spectrum was calibrated with sulfur dioxide (SO$_2$) rotational 
lines whose spectral frequencies are well known \citep{SO2_rot_2005}. A portion of the 
source radiation propagated through a Fabry-Perot cavity to produce an interference fringe 
spectrum with a free spectral range of $\sim$9.2~MHz. The frequencies of radiation between 
the calibration lines were interpolated with the fringe spectrum. 
In the calibration procedure, the dispersive effect of atmospheric water vapor in the 
Fabry-Perot cavity was taken into account \citep{dispersion-corr_2006,dispersion-corr_2007}. 
Measurements were taken with scans that proceeded both upward and downward in frequency 
so as to record an average frequency. The results obtained from 100 upward and downward 
scans were accumulated for a better signal-to-noise ratio, increasing the integration 
time from $\sim$0.1 to $\sim$10~ms. The experimental uncertainty of this apparatus is 
around 50~kHz for an isolated, well-calibrated line.

We used phase-lock loop (PLL) systems in the Cologne spectrometers to obtain accurate 
frequencies. Two BWOs were used as sources to record lines in the 566$-$670 and 
848$-$930~GHz regions. A portion of the radiation from the BWOs was mixed with an 
appropriate harmonic of a continuously tunable synthesizer in a Schottky diode 
multiplier mixer to produce the intermediate frequency (IF) signal. The IF-signal 
was phase-locked and phase-error provided by the PLL circuit were fed back to the power 
supply of the BWOs. The experimental uncertainties under normal absorption conditions 
can be as low as 5 kHz even around 1~THz \citep{CO-BWO_1995,SiS_rot_2007}.

A solid-state based spectrometer system was used to obtain transition frequencies between 
1290 and 1390~GHz. A set of frequency multipliers (three doublers plus two triplers) were 
driven by a microwave synthesizer to cover these frequencies. Additional detail is given 
by \citet{CH3SH_rot_THz-chain_2012}. Accuracies of 10~kHz can be achieved for strong, 
isolated lines \citep{MeCN_v8_le_2_rot_2015,H2CO-X_rot_2017}.

Thioformaldehyde (H$_2$CS) was produced by the pyrolysis of trimethylene sulfide 
[(CH$_2$)$_3$S; Sigma-Aldrich Co.], which was used as provided. The thermal 
decomposition of trimethylene sulfide yields mostly thioformaldehyde and ethylene, 
which has no permanent dipole moment, but also small amounts of by-products such 
as CS, H$_2$S, and H$_2$CCS. Laboratory setups for the pyrolysis were slightly 
different in the OSU and Cologne measurements. At OSU, trimethylene sulfide vapor 
was passed through a 2~cm diameter, 20~cm long piece of quartz tubing stuffed with 
quartz pieces and quartz cotton to enlarge the reaction surface. The quartz tubing 
was heated with a cylindrical furnace to $\sim$680$^{\circ}$C. The gas produced from 
the pyrolysis was introduced to a 6~m long aluminum cell at room temperature and 
pumped to a pressure of 0.4$-$1.5~mTorr (1~mTorr = 0.1333~Pa). The spectrum of 
trimethylene sulfide almost totally disappeared after the pyrolysis, at which time 
the spectrum of thioformaldehyde appeared. Spectral lines due to the by-products, 
CS, H$_2$S, and H$_2$CCS, were also observed, but with less intensity compared 
with thioformaldehyde.

A 3~m long glass absorption cell kept at room temperature was used for measurements 
in Cologne. A higher temperature of about 1300$^{\circ}$C was required in the pyrolysis 
zone in order to maximize the thioformaldehyde yield and to minimize absorptions 
of (CH$_2$)$_3$S because no quartz cotton was used in the quartz pyrolysis tube. 
The total pressure was around 1$-$3~Pa for weaker lines and around 0.01$-$0.1~Pa 
for stronger lines.

Liquid He-cooled InSb bolometers were used in both laboratories as detectors.

\section{Quantum-chemical calculations}
\label{qcc}

Hybrid density functional calculations of the B3LYP variant \citep{Becke_1993,LYP_1988} 
and M{\o}ller-Plesset second order perturbation theory (MP2) calculations \citep{MPn_1934} 
were carried out with the commercially available program Gaussian~09 \citep{G09E}. 
We performed also coupled cluster calculations with singles and doubles excitations 
augmented by a perturbative correction for triple excitations, CCSD(T) \citep{CC+T_1989} 
with the 2005 Mainz-Austin-Budapest version of ACESII and its successor 
CFOUR\footnote{CFOUR, a quantum chemical program package written by J.~F. Stanton, J. Gauss,
M.~E. Harding, P.~G. Szalay et al. For the current version, see http://www.cfour.de}. 
We employed correlation consistent basis sets cc-pVXZ (X = T, Q, 5) \citep{cc-pVXZ_1989} 
for H and C and the cc-pV(X+d)Z basis sets for S \citep{plusD_1995}; diffuse basis 
functions were augmented for some calculations, denoted as aug-cc-pVXZ and aug-cc-pV(X+d)Z. 
We abbreviate these basis sets as XZ and aXZ, respectively. In addition, we employed 
weighted core-correlating basis functions in some cases, yielding the (aug-) cc-pwCVXZ 
basis sets \citep{core-corr_2002}. These basis sets were abbreviated as wCXZ and awCXZ, 
respectively. All calculation were carried aut at the Regionales Rechenzentrum der 
Universit{\"a}t zu K{\"o}ln (RRZK).

Equilibrium geometries were determined by analytic gradient techniques, harmonic 
force fields by analytic second derivatives, and anharmonic force fields by 
numerical differentiation of the analytically evaluated second derivatives of 
the energy. The main goal of these anharmonic force field calculations was 
to evaluate first order vibration-rotation parameters \citep{vib-rot_rev_1972}, 
see also Sect.~\ref{structure}. Core electrons were kept frozen in MP2 and CCSD(T) 
calculations unless ``ae'' indicates that all electrons were correlated. We evaluated 
the hyperfine parameters of H$_2$C$^{33}$S using the awCQZ basis set (wCQZ for CCSD(T) 
calculation) at the equilibrium geometry calculated at the same level.

\section{Spectroscopic properties of thioformaldehyde}
\label{rot_vib_backgr}

Thioformaldehyde is an asymmetric rotor with $\kappa = (2B - A - C)/(A - C) = -0.9924$, 
much closer to the symmetric limit of $-1$ than the isovalent formaldehyde for which 
$\kappa$ is $-$0.9610, see, for example, \citet{H2CO-X_rot_2017}. 
The H$_2$CS dipole moment of 1.6491~D \citep{H2CS_dip_1977} is aligned with the $a$ 
inertial axis. The strong rotational transitions are therefore those with $\Delta K_a = 0$ 
and $\Delta J = +1$, that is, the \textit{R}-branch transitions. Transitions with $\Delta K_a = 0$ 
and $\Delta J = 0$ (\textit{Q}-branch transitions) are also allowed as are transitions 
with $\Delta K_a = \pm2$. These transitions are not only much weaker than the strong 
\textit{R}-branch transitions, but also relatively weaker than the equivalent transitions 
in H$_2$CO because H$_2$CS is closer to the symmetric prolate limit.

Isotopologs with two H or two D have $C_{\rm 2v}$ symmetry whereas isotopologs with one H 
and one D have $C_{\rm S}$ symmetry. Spin-statistics caused by the two equivalent H lead 
to \textit{ortho} and \textit{para} states with a 3~:~1 intensity ratio. The \textit{ortho} 
states are described by $K_a$ being odd. The \textit{ortho} to \textit{para} ratio in 
D$_2$CS is 2~:~1, and the \textit{ortho} states are described by $K_a$ being even. 
No non-trivial spin-statistics exist in HDCS and related isotopologs.

Sulfur has four stable isotopes with mass numbers 32, 33, 34, and 36 and with terrestrial 
abundances of 95.0\,\%, 0.75\,\%, 4.2\,\%, and $\sim$0.015\,\%, respectively 
\citep{iso-comp_2011}. The respective abundances are 98.89\,\% and 1.11\,\% for $^{12}$C 
and $^{13}$C and 99.98\,\% and $\sim$0.015\,\% for H and D.

\section{Spectroscopic results}
\label{lab-results}

We used Pickett's SPCAT and SPFIT programs \citep{spfit_1991} to predict and fit rotational 
spectra of the various isotopic species of thioformaldehyde. Predictions were generated 
from the published data for the isotopic species H$_2$CS, H$_2$C$^{34}$S, H$_2$C$^{33}$S, 
H$_2^{13}$CS, and HDCS 
\citep{H2CS_isos_rot_1971,H2CS_rot_1972,H2CS_isos_1982,H2CS_HFS_1987,H2CS_FIR_nu2_1993,HDCS_rot_det_1997}. 
Higher order spectroscopic parameters of isotopic species with heavy atom substition were 
estimated from those of H$_2$CS by scaling the parameters with appropriate powers of the ratios 
of $2A - B - C$, $B + C$, and $B - C$. Even though these estimates do not hold strictly, they 
are almost always better than constraining the parameters to zero and also mostly better than 
contraining the parameters to values directly taken from the main isotopic species, see, 
e.g., below or the examples of isotopic CH$_3$CN \citep{MeCN_13C-vib_rot_2016} or H$_2$CO 
\citep{H2CO-X_rot_2017}. This scaling procedure is not recommended for H to D substitution 
especially in molecules with relatively few atoms, such as HDCS. 
The resulting spectroscopic line lists are or were available in the Cologne Database for 
Molecular Spectroscopy, CDMS\footnote{https://cdms.astro.uni-koeln.de/classic/entries/} 
\citep{CDMS_3} as version 1, mostly from February 2006. An updated entry (version 2) 
has been available for the H$_2$CS main isotopic species since early 2008.


\begin{figure}
\centering
\includegraphics[width=8cm,angle=0]{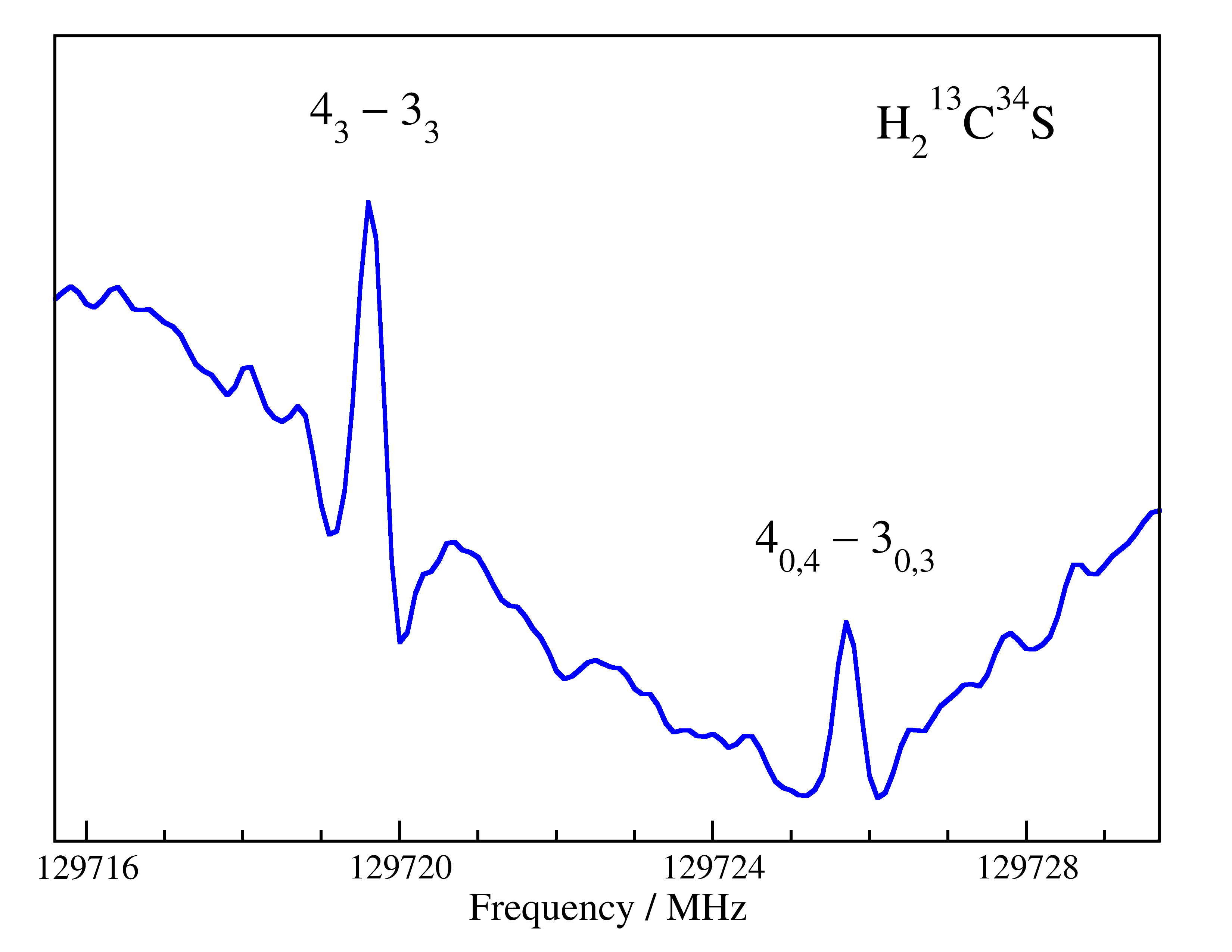}

\caption{Section of the rotational spectrum of H$_2^{13}$C$^{34}$S recorded in natural isotopic 
   composition, showing two low-$J$ lines. $K_c$ was omitted for the $K_a = 3$ transitions 
   because the asymmetry splitting was not resolved and $K_c = J - K_a$ or $K_c = J - K_a + 1$.}
\label{1334_fig}
\end{figure}


We carried out the rotational assignment for the FASSST spectra of thioformaldehyde with 
the Computer Aided Assignment of Asymmetric Rotor Spectra (CAAARS) program applying the 
Loomis-Wood procedure, with which the observed spectrum is visually compared with predicted 
line positions and intensities to make new assignments \citep{CAAARS_2005}. 
The strong $\Delta K_a = 0$ \textit{R}-branch transitions of the abundant isotopic species 
H$_2$CS, H$_2$C$^{34}$S, H$_2^{13}$CS, and H$_2$C$^{33}$S were found easily first for low 
values of $K_a$ (0$-$4) and later up to $K_a = 9$. The upper frequency of 377~GHz limited 
the $J$ quantum numbers to a maximum of $10 - 9$ for H$_2$CS and $11 - 10$ for the other 
isotopologs because of the smaller values of $B + C$. Several transitions of H$_2$C$^{33}$S 
displayed splitting caused by the electric nuclear quadrupole moment and the magnetic 
nuclear dipole moment of the $I = 3/2$ nucleus of $^{33}$S. We could also make extensive 
assignments for the weaker \textit{Q}-branch transitions with $K_a = 1$. These covered 
all and almost all of $J = 15$ to 26 for H$_2$CS and H$_2$C$^{34}$S, respectively; fewer 
lines were found for H$_2^{13}$CS and H$_2$C$^{33}$S. In the case of the main isotopic 
species, we could assign most of the even weaker $K_a = 2$ \textit{Q}-branch transitions 
with $32 \le J \le 41$.


\begin{figure}
\centering
\includegraphics[width=8cm,angle=0]{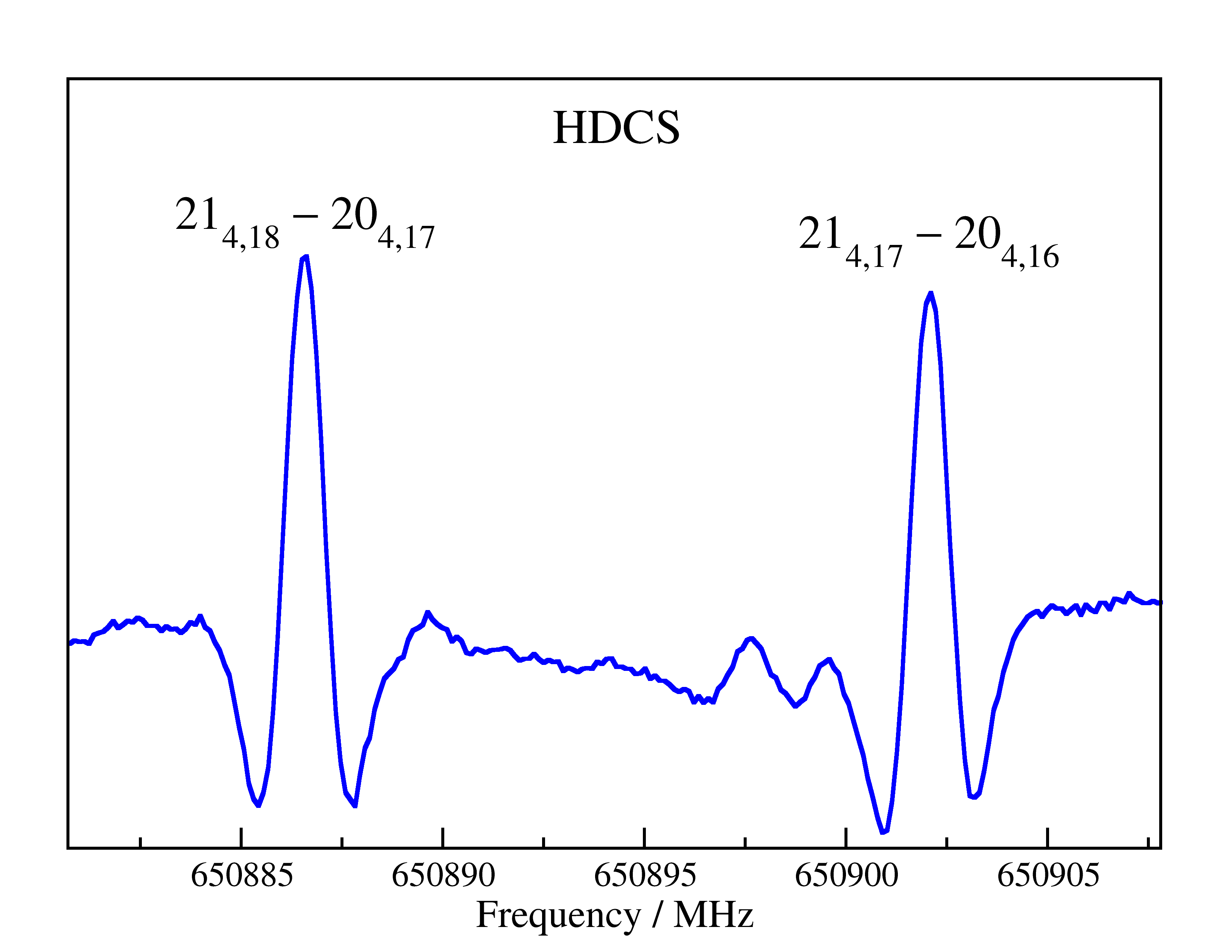}

\caption{Section of the rotational spectrum of HDCS recorded in natural isotopic composition, 
   displaying the $K_a = 4$ asymmetry splitting. The identities of the weaker features are not known.}
\label{HDCS_K=4_fig}
\end{figure}


\begin{figure}
\centering
\includegraphics[width=8cm,angle=0]{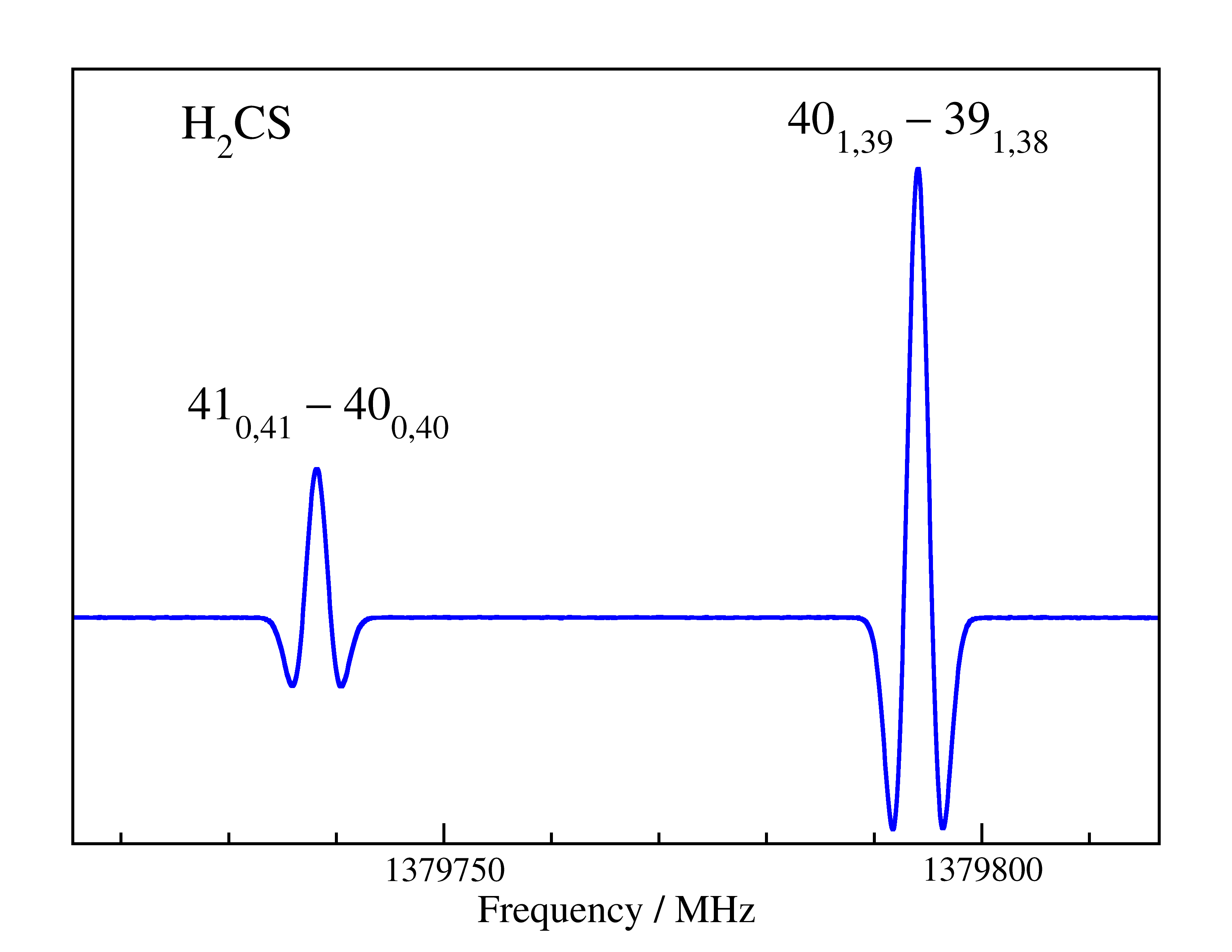}

\caption{Section of the rotational spectrum of H$_2$CS displaying the \textit{para} to 
\textit{ortho} ratio of 1 to 3.}
\label{H2CS_opr_fig}
\end{figure}

On the basis of these extensive assignments, we suspected that transitions of HDCS should 
be strong enough in natural isotopic composition to identify them in our FASSST spectra. 
\citet{HDCS_rot_det_1997} had reported transition frequencies from laboratory measurements 
up to 380~GHz. The resulting spectroscopic parameters were sufficiently accurate to 
identify HDCS transitions in our FASSST spectra. Even though the lines were weak, 
we could supplement the existing line list with \textit{R}-branch transition frequencies 
not reported by \citet{HDCS_rot_det_1997} up to $K_a = 7$.


\begin{table*}
\begin{center}
\caption{Spectroscopic parameters$^{a,b}$ (MHz) of thioformaldehyde isotopologs with different sulfur isotopes.}
\label{tab-parameters_S32-36}
\renewcommand{\arraystretch}{1.10}
\begin{tabular}[t]{lr@{}lr@{}lr@{}lr@{}l}
\hline \hline
Parameter & \multicolumn{2}{c}{H$_2$CS} & \multicolumn{2}{c}{H$_2$C$^{33}$S} & \multicolumn{2}{c}{H$_2$C$^{34}$S} & \multicolumn{2}{c}{H$_2$C$^{36}$S} \\
\hline
$A- (B+C)/2$             & 274437&.5932~(115)    & 274588&.054~(306)    & 274729&.12~(34)      & 274987&.91~(94)     \\
$(B+C)/2$                &  17175&.745955~(196)  &  17024&.740821~(110) &  16882&.911552~(112) &  16621&.73726~(35)  \\
$(B-C)/4$                &    261&.6240523~(165) &    257&.050936~(39)  &    252&.793027~(73)  &    245&.04552~(43)  \\
$D_K$                    &     23&.34378~(164)   &     23&.408          &     23&.468~(141)    &     23&.6           \\
$D_{JK}$                 &      0&.5222938~(43)  &      0&.5132638~(95) &      0&.5048431~(51) &      0&.489486~(38) \\
$D_J \times 10^3$        &     19&.01875~(39)    &     18&.700456~(105) &     18&.404173~(172) &     17&.86334~(29)  \\
$d_1 \times 10^3$        &   $-$1&.208429~(105)  &   $-$1&.176656~(78)  &   $-$1&.148425~(108) &   $-$1&.09806~(36)  \\
$d_2 \times 10^3$        &   $-$0&.1773270~(222) &   $-$0&.171180~(81)  &   $-$0&.165589~(136) &   $-$0&.15585~(27)  \\
$H_K \times 10^3$        &      5&.946~(35)      &      5&.97           &      6&.00           &      6&.05          \\
$H_{KJ} \times 10^6$     &  $-$28&.155~(86)      &  $-$27&.839~(211)    &  $-$28&.071~(109)    &  $-$28&.16~(61)     \\
$H_{JK} \times 10^6$     &      1&.50409~(270)   &      1&.4502~(39)    &      1&.41629~(70)   &      1&.3346~(203)  \\
$H_J \times 10^9$        &   $-$5&.81~(32)       &   $-$5&.46           &   $-$5&.100~(40)     &   $-$4&.48          \\
$h_1 \times 10^9$        &      3&.018~(141)     &      2&.792          &      2&.600~(37)     &      2&.216         \\
$h_2 \times 10^9$        &      1&.6472~(140)    &      1&.524          &      1&.415~(49)     &      1&.209         \\
$h_3 \times 10^9$        &      0&.3619~(73)     &      0&.3393         &      0&.3186~(144)   &      0&.2796        \\
$L_K \times 10^6$        &   $-$2&.109~(206)     &   $-$2&.00           &   $-$2&.00           &   $-$2&.00          \\
$L_{KKJ} \times 10^9$    &  $-$21&.36~(69)       &  $-$23&.18~(128)     &  $-$20&.86~(65)      &  $-$20&.37          \\
$L_{JK} \times 10^9$     &      0&.2032~(90)     &      0&.200          &      0&.197          &      0&.1909        \\
$L_{JJK} \times 10^{12}$ &  $-$10&.32~(81)       &   $-$9&.66           &   $-$9&.0            &   $-$7&.85          \\
$L_J \times 10^{12}$     &      0&.833~(87)      &      0&.766          &      0&.700          &      0&.588         \\
$l_1 \times 10^{12}$     &   $-$0&.358~(47)      &   $-$0&.330          &   $-$0&.304          &   $-$0&.258         \\
$P_{KKJ} \times 10^{12}$ &  $-$18&.63~(180)      &  $-$18&.8            &  $-$19&.0            &  $-$19&.0           \\
\hline
\end{tabular}
\end{center}
\tablefoot{
$^{(a)}$ Watson's $S$ reduction has been used in the representation $I^r$. Numbers in parentheses are one standard 
         deviation in units of the least significant figures. Parameters without uncertainties were estimated and 
         kept fixed in the analyses.
$^{(b)}$ $^{33}$S HFS parameters are given in Table~\ref{tab-HFS-parameters}. 
}
\end{table*}


\begin{table*}
\begin{center}
\caption{Experimental $^{33}$S hyperfine structure parameters$^a$ (MHz) of H$_2$C$^{33}$S in comparison 
         to equilibrium values from quantum-chemical calculations$^b$.}
\label{tab-HFS-parameters}
\renewcommand{\arraystretch}{1.10}
\begin{tabular}[t]{lr@{}lr@{}lr@{}lr@{}lr@{}l}
\hline \hline
Parameter & \multicolumn{2}{c}{exptl.} & \multicolumn{2}{c}{B3LYP} & \multicolumn{2}{c}{MP2} & \multicolumn{2}{c}{ae-MP2} & \multicolumn{2}{c}{ae-CCSD(T)} \\
\hline
$\chi _{aa}$                             & $-$11&.8893~(124) & $-$12&.27 & $-$10&.31 &  $-$9&.99 & $-$12&.818 \\
$\chi _{bb}$                             &    49&.9668~(156) &    50&.26 &    49&.08 &    48&.93 &    50&.030 \\
$\chi _{cc}$$^c$                         & $-$38&.0775~(158) & $-$37&.99 & $-$38&.77 & $-$38&.94 & $-$37&.212 \\
$(C_{aa}-(C_{bb}+C_{cc})/2) \times 10^3$ &   475&.5~(24)     &   526&.5  &   469&.3  &   468&.5  &   456&.0   \\
$(C_{bb} + C_{cc}) \times 10^3$          &    13&.6          &    15&.28 &    13&.61 &    13&.63 &    13&.6   \\
$(C_{bb} - C_{cc}) \times 10^3$          &    10&.68~(105)   &    10&.98 &    10&.29 &    10&.33 &    10&.2   \\
\hline
\end{tabular}
\end{center}
\tablefoot{
$^{(a)}$ Numbers in parentheses are one standard deviation in units of the least significant figures. 
         Parameters without uncertainties were estimated and kept fixed in the analyses.
$^{(b)}$ Basis sets: aug-cc-pwCVQZ for B3LYP and MP2 calculations, cc-pwCVQZ for ae-CCSD(T); see also section~\ref{qcc}. 
$^{(c)}$ Derived value because the sum of the $\chi _{ii}$ is zero.
}
\end{table*}


\begin{table*}
\begin{center}
\caption{Spectroscopic parameters$^a$ (MHz) of H$_2^{13}$CS, H$_2^{13}$C$^{34}$S, HDCS, and D$_2$CS.}
\label{tab-parameters_13C_D}
\renewcommand{\arraystretch}{1.10}
\begin{tabular}[t]{lr@{}lr@{}lr@{}lr@{}l}
\hline \hline
Parameter & \multicolumn{2}{c}{H$_2^{13}$CS} & \multicolumn{2}{c}{H$_2^{13}$C$^{34}$S} & \multicolumn{2}{c}{HDCS} & \multicolumn{2}{c}{D$_2$CS}\\
\hline
$A- (B+C)/2$             & 275113&.82~(35)      & 275418&.9~(161)     & 187214&.14~(44)       & 132198&.92~(26)      \\
$(B+C)/2$                &  16514&.989247~(141) &  16219&.21918~(122) &  15501&.179704~(242)  &  14200&.0562~(59)    \\
$(B-C)/4$                &    241&.898578~(82)  &    233&.32414~(81)  &    304&.81583~(44)    &    352&.105440~(153) \\
$D_K$                    &     23&.465~(129)    &     23&.51          &     13&.364~(198)     &      5&.5            \\
$D_{JK}$                 &      0&.4960263~(70) &      0&.478799~(57) &      0&.3208611~(107) &      0&.29091~(146)  \\
$D_J \times 10^3$        &     17&.692846~(203) &     17&.0998~(66)   &     15&.40446~(68)    &     12&.492~(158)    \\
$d_1 \times 10^3$        &   $-$1&.071785~(122) &   $-$1&.01853       &   $-$1&.38941~(86)    &   $-$1&.40268~(225)  \\
$d_2 \times 10^3$        &   $-$0&.152488~(113) &   $-$0&.14239       &   $-$0&.23009~(42)    &   $-$0&.28995~(37)   \\
$H_K \times 10^3$        &      6&.00           &      6&.00          &      3&.0             &      0&.75           \\
$H_{KJ} \times 10^6$     &  $-$25&.846~(164)    &  $-$25&.77          &  $-$37&.662~(110)     &   $-$4&.7            \\
$H_{JK} \times 10^6$     &      1&.34969~(281)  &      1&.271         &      1&.2332~(61)     &      0&.88           \\
$H_J \times 10^9$        &   $-$5&.660~(45)     &   $-$4&.97          &      1&.885~(288)     &      1&.3            \\
$h_1 \times 10^9$        &      2&.577~(39)     &      2&.21          &      2&.75~(52)       &      3&.0            \\
$h_2 \times 10^9$        &      1&.292~(38)     &      1&.11          &      1&.92~(32)       &      1&.719~(109)    \\
$h_3 \times 10^9$        &      0&.3009~(110)   &      0&.2645        &      0&.50            &      0&.792~(33)     \\
$L_K \times 10^6$        &   $-$2&.00           &   $-$2&.00          &       &               &       &              \\
$L_{KKJ} \times 10^9$    &  $-$22&.16~(106)     &  $-$21&.65          &       &               &       &              \\
$L_{JK} \times 10^9$     &      0&.189          &      0&.183         &       &               &       &              \\
$L_{JJK} \times 10^{12}$ &  $-$10&.06           &   $-$8&.80          &       &               &       &              \\
$L_J \times 10^{12}$     &      0&.750          &      0&.630         &       &               &       &              \\
$l_1 \times 10^{12}$     &   $-$0&.320          &   $-$0&.272         &       &               &       &              \\
$P_{KKJ} \times 10^{12}$ &  $-$19&.0            &  $-$19&.0           &       &               &       &              \\
\hline
\end{tabular}
\end{center}
\tablefoot{
$^{(a)}$ Watson's $S$ reduction has been used in the representation $I^r$. Numbers in parentheses are one standard 
         deviation in units of the least significant figures. Parameters without uncertainties were estimated and 
         kept fixed in the analyses.}
\end{table*}


With the identification of HDCS in the FASSST spectra, it appeared plausible to search 
for transitions of H$_2^{13}$C$^{34}$S and H$_2$C$^{36}$S because they have abundances 
similar to those of HDCS. We derived an $r_{I,\epsilon}$ structure \citep{r_I-eps_1991}, 
see also Sect.~\ref{structure}, from the known rotational parameters because this structure 
model can provide good predictions of rotational parameters of isotopologs not yet studied, 
see, e.g., the example of cyclopropylgermane \citep{use_r_I-eps_c-PrGeH3_1992} or sulfuryl 
chloride fluoride \citep{use_r_I-eps_SO2ClF_1994}. We were able to make assignments of 
\textit{R}-branch transition up to $K_a = 5$ for both isotopic species, however, those for 
H$_2^{13}$C$^{34}$S were only made in the process of writing this manuscript, and assignments 
extend only up to 362~GHz.

Most of the frequencies were assigned uncertainties of 50~kHz; 100~kHz were assigned to 
the weak \textit{Q}-branch transition frequencies, to some of the weaker H$_2$C$^{33}$S 
lines, and to the stronger H$_2$C$^{36}$S lines. Uncertainties of 200~kHz were assigned 
to weaker lines of H$_2$C$^{36}$S.

Subsequently, improved predictions were used to search for individual transitions 
of H$_2$CS, H$_2$C$^{34}$S, H$_2^{13}$CS, H$_2$C$^{33}$S, HDCS, and H$_2$C$^{36}$S 
in the regions 566$-$670 and 848$-$930~GHz using the Cologne Terahertz Spectrometer. 
Fig.~\ref{HDCS_K=4_fig} demonstrates the good signal-to-noise ratio achieved for HDCS. 
Later, we recorded transitions of H$_2$CS, H$_2$C$^{34}$S, and H$_2^{13}$CS in the 
1290$-$1390~GHz region. Fig.~\ref{H2CS_opr_fig} demonstrates the \textit{para} to 
\textit{ortho} ratio for the main isotopic species. The Boltzmann peak of the room 
temperature rotational spectrum of H$_2$CS is at $\sim$800~GHz. Therefore, we did not 
attempt any measurements for the rarer isotopic species at these high frequencies.

The quantum numbers of the strong $R$-branch transitions reach $J = 41 - 40$ and 
$K_a = 15$ for H$_2$CS; we recorded three $K_a = 1$ \textit{Q}-branch transitions 
up to $J = 41$ and five $K_a = 2 - 0$ or $3 - 1$ \textit{P}-branch transitions 
($\Delta J = -1$) below 1~THz. In the case of H$_2$C$^{34}$S and H$_2^{13}$CS, 
$J = 42 - 41$, $K_a = 12$ and $J = 43 - 42$, $K_a = 11$ were reached. In addition, 
we recorded two $K_a = 1$ \textit{Q}-branch transitions with $J = 34$ and 35 for 
H$_2$C$^{34}$S. Finally, $J = 27 - 26$, $K_a = 11$ and $J = 28 - 27$, $K_a = 7$, 
and $J = 30 - 29$, $K_a = 9$ were reached for H$_2$C$^{33}$S, H$_2$C$^{36}$S, 
and HDCS, respectively. Unfortunately, we did not attempt to search for transitions 
of H$_2^{13}$C$^{34}$S in the (initial) absence of assignments in the FASSST spectra.

Uncertainties of 5~kHz were assigned to the best lines of almost all isotopic species 
below 1~THz, 10~kHz were assigned to the best lines of H$_2$C$^{36}$S and those above 
1~THz. The largest uncertainties were around 50~kHz.

Our data set for the main isotopic species is very similar to that of our earlier account 
by \citet{H2CS_rot_2008}. The main exception are 58 transitions corresponding to 42 
frequencies because of unresolved asymmetry splitting at higher $K_a$ that were recorded 
between 1290 and 1390~GHz. We omitted the far-infrared transition frequencies from 
\citet{H2CS_FIR_nu2_1993} below 1390~GHz or 46.37~cm$^{-1}$ because of the lower accuracy 
of these data ($\sim$3~MHz). Additional data beyond our transition frequencies are the 
ground state combination differences from the A$-$X electronic spectrum of H$_2$CS 
\citep{H2CS_A-X_1994} and lower frequency rotational transitions frequencies 
\citep{H2CS_dip_1977,H2CS_isos_rot_1971,H2CS_rot_1972}. Some transition frequencies 
for H$_2$C$^{34}$S and H$_2^{13}$CS were taken from \citet{H2CS_isos_rot_1971} and 
from \citet{H2CS_isos_1982}; \citet{H2CS_HFS_1987} contributed data for H$_2$C$^{33}$S 
and for H$_2^{13}$CS, and \citet{HDCS_rot_det_1997} for HDCS. We determined also 
spectroscopic parameters for D$_2$CS in particular for the structure determination 
even though we did not record any transitions of this rare isotopolog. We combined 
earlier laboratory data \citep{H2CS_isos_rot_1971,H2CS_isos_1982} with more recent 
rest frequencies from radio astronomical observations \citep{D2CS_det_2005}.

The HFS components of the $2_{1,1} - 2_{1,2}$ transition of H$_2$C$^{33}$S 
\citep{H2CS_HFS_1987} displayed average residuals between measured and calculated 
frequencies of 28~kHz, much larger than the assigned uncertainties of 4$-$8~kHz 
and with considerable scatter. Therefore, we omitted the HFS components of this 
transition from the final fit.

The additional very acurate data for the main isotopic species from the 1290$-$1390~GHz 
region required two additional parameters, $L_{JJK}$ and $L_J$, in the fit compared with 
our previous report \citep{H2CS_rot_2008}. The parameter values changed only very little 
for the most part with the exception of $H_J$, which changed from $(-3.33 \pm 0.29)$~mHz 
to $(-5.81 \pm 0.31)$~mHz.

Sets of spectroscopic parameters were evaluated for H$_2$C$^{34}$S and H$_2^{13}$CS as 
described above. Parameters derived from the main isotopic species were fit starting 
from the lower order parameters. A parameter was considered to be kept floating if 
the resulting quality of the fit improved substantially and if the uncertainty of 
the parameter was much smaller than a fifth of the magnitude of the parameter. 
Generally, we searched for the parameter whose fitting improved the quality most among 
the parameters reasonable to be fit. This procedure was repeated until the quality of 
the fit did not improve considerably anymore.

We evaluated initial spectroscopic parameters of H$_2$C$^{33}$S, H$_2$C$^{36}$S, and 
H$_2^{13}$C$^{34}$S in a similar manner; the main difference was that we considered 
not only parameters of H$_2$CS, but also of H$_2$C$^{34}$S and H$_2^{13}$CS. 
Nuclear HFS parameters had to be included in the fit of H$_2$C$^{33}$S. The dominant 
contribution comes from the nuclear electric quadrupole coupling. There are only three 
parameters, $\chi_{aa}$, $\chi_{bb}$, and $\chi_{cc}$, because of the symmetry of 
the molecule; $\chi_{cc}$ was derived from the other two because the sum of the three 
is zero. Nuclear magnetic spin-rotation coupling parameters needed to be included 
in the fit also. However, the values of $C_{bb}$ and $C_{cc}$ were poorly determined 
and were quite different from values obtained from quantum-chemical calculations. 
It turned out that $C_{bb} - C_{cc}$ is well determined whereas $C_{bb} + C_{cc}$ 
appears to be insufficiently constrained. Therefore, we constrained $C_{bb} + C_{cc}$ 
to the value from an MP2 calculation because the remaining two experimental parameters 
agreed well with the calculated ones.
Distortion parameters of HDCS and D$_2$CS that could not be evaluated experimentally were 
estimated from values taken from a quantum-chemical calculation \citep{H2CS_FF_ai_1994} 
and considering deviations between these calculated equilibrium values and the determined 
experimental ground state values of these two isotopic species and those of H$_2$CS.


\begin{table}
\begin{center}
\caption{Frequencies $\nu$ (GHz) and $J$ values for which deviations between old calculations 
         and present data exceed 1~MHz given for selected thioformaldehyde isotopic species 
         and $K_a$ values.}
\label{deviations}
\renewcommand{\arraystretch}{1.10}
\begin{tabular}[t]{lccr}
\hline \hline
Isotopolog     & $K_a$ & $\nu$     & $J$      \\
\hline
H$_2^{13}$CS   & 5     & $> 198.0$ & $\ge 5$  \\
               & 2u$^a$& $> 330.5$ & $\ge 9$  \\
               & 0     & $> 461.0$ & $\ge 13$ \\
H$_2$C$^{34}$S & 5     & $> 202.4$ & $\ge 5$  \\
               & 4     & $> 269.9$ & $\ge 7$  \\
               & 3     & $> 405.0$ & $\ge 11$ \\
H$_2$C$^{33}$S & 6     & $> 238.0$ & $\ge 6$  \\
               & 5     & $> 306.1$ & $\ge 8$  \\
               & 4     & $> 476.3$ & $\ge 13$ \\
\hline
\end{tabular}
\end{center}
\tablefoot{
$^{(a)}$ Transitions with $K_c = J - 2$; transitions with 
  $K_c = J - 1$ deviate less. 
}
\end{table}


The spectroscopic parameters of H$_2$CS, H$_2$C$^{33}$S, H$_2$C$^{34}$S, and H$_2$C$^{36}$S 
are given in Table~\ref{tab-parameters_S32-36}, except for the H$_2$C$^{33}$S hyperfine 
structure parameters which are presented in Table~\ref{tab-HFS-parameters} in comparison to 
values from quantum-chemical calculations. The spectroscopic parameters of H$_2^{13}$CS, 
H$_2^{13}$C$^{34}$S, HDCS, and D$_2$CS are gathered in Table~\ref{tab-parameters_13C_D}.

The experimental data have been reproduced within experimental uncertainties for all isotopic 
species. There is some scatter among partial data, e.g, previous rotational lines for a given 
isotopolog. Some partial data sets have been judged slightly conservatively.


\begin{table*}
\begin{center}
\caption{Spectroscopic parameters$^a$ (MHz) of thioformaldehyde in comparison to those of related molecules.}
\label{tab_comp-parameters}
\renewcommand{\arraystretch}{1.10}
\begin{tabular}[t]{lr@{}lr@{}lr@{}lr@{}l}
\hline \hline
Parameter & \multicolumn{2}{c}{H$_2$CO$^b$} & \multicolumn{2}{c}{H$_2$CS$^c$} & \multicolumn{2}{c}{H$_2$SiO$^d$} & \multicolumn{2}{c}{H$_2$SiS$^e$} \\
\hline
$A- (B+C)/2$             & 245551&.4495~(40)    & 274437&.5932~(115)     & 148946&.49~(173)    & 162498&.0~(14)      \\
$(B+C)/2$                &  36419&.11528~(25)   &  17175&.745955~(196)   &  17711&.07958~(33)  &   7844&.48028~(22)  \\
$(B-C)/4$                &   1207&.4358721~(33) &    261&.6240523~(165)  &    483&.74088~(50)  &     93&.23731~(28)  \\
$D_K$                    &     19&.39136~(53)   &     23&.34378~(164)    &      7&.63~(87)     &      9&.811         \\
$D_{JK}$                 &      1&.3211073~(93) &      0&.5222938~(43)   &      0&.610532~(60) &      0&.151376~(44) \\
$D_J \times 10^3$        &     70&.32050~(50)   &     19&.01875~(39)     &     16&.1803~(94)   &      3&.92823~(27)  \\
$d_1 \times 10^3$        &  $-$10&.437877~(47)  &   $-$1&.208429~(105)   &   $-$2&.08116~(236) &   $-$0&.19581~(35)  \\
$d_2 \times 10^3$        &   $-$2&.501496~(33)  &   $-$0&.1773270~(222)  &   $-$0&.6712~(48)   &   $-$0&.02938~(17)  \\
$H_K \times 10^3$        &      4&.027~(22)     &      5&.946~(35)       &      1&.0$^f$       &      1&.6$^f$       \\
$H_{KJ} \times 10^6$     &     10&.865~(79)     &  $-$28&.155~(86)       &  $-$43&.324~(297)   &  $-$18&.8~(15)      \\
$H_{JK} \times 10^6$     &      7&.465~(16)     &      1&.50409~(270)    &      3&.409~(65)    &      0&.246~(34)    \\
$H_J \times 10^9$        &      3&.54~(33)      &   $-$5&.81~(32)        &       &             &       &             \\
$h_1 \times 10^9$        &     32&.272~(58)     &      3&.018~(141)      &       &             &       &             \\
$h_2 \times 10^9$        &     47&.942~(74)     &      1&.6472~(140)     &       &             &       &             \\
$h_3 \times 10^9$        &     15&.966~(15)     &      0&.3619~(73)      &       &             &       &             \\
$L_K \times 10^6$        &   $-$0&.610~(177)    &   $-$2&.109~(206)      &       &             &       &             \\
$L_{KKJ} \times 10^9$    &   $-$5&.85~(19)      &  $-$21&.36~(69)        &       &             &       &             \\
$L_{JK} \times 10^9$     &      0&.367~(85)     &      0&.2032~(90)      &       &             &       &             \\
$L_{JJK} \times 10^{12}$ & $-$105&.7~(92)       &  $-$10&.32~(81)        &       &             &       &             \\
$L_J \times 10^{12}$     &       &              &      0&.833~(87)       &       &             &       &             \\
$l_1 \times 10^{12}$     &       &              &   $-$0&.358~(47)       &       &             &       &             \\
$l_2 \times 10^{12}$     &   $-$0&.345(50)      &       &                &       &             &       &             \\
$l_3 \times 10^{12}$     &   $-$0&.427(19)      &       &                &       &             &       &             \\
$l_4 \times 10^{12}$     &   $-$0&.1520~(32)    &       &                &       &             &       &             \\
$P_{KKJ} \times 10^{12}$ &       &              &  $-$18&.63~(180)       &       &             &       &             \\
$p_5 \times 10^{18}$     &      3&.33           &       &                &       &             &       &             \\
\hline
\end{tabular}
\end{center}
\tablefoot{
$^{(a)}$ Watson's $S$ reduction has been used in the representation $I^r$. Numbers in parentheses are one standard 
         deviation in units of the least significant figures. Parameters without uncertainties were estimated and 
         kept fixed in the analyses.
$^{(b)}$ \citet{H2CO-X_rot_2017}.
$^{(c)}$ This work. 
$^{(d)}$ \citet{H2SiO_rot_1994}; refit in the $S$ reduction in the present work.
$^{(e)}$ \citet{H2SiS_Si2S_rot_2011}
$^{(f)}$ Estimated in the present work.
}
\end{table*}


The experimental transition frequencies with quantum numbers, uncertainties, and differences
to the calculated frequencies in the final fits are available as supplementary material to 
this article. The line, parameter, and fit files along with auxiliary files are available 
in the data section of the CDMS.\footnote{https://cdms.astro.uni-koeln.de/classic/predictions/daten/H2CS/}
Calculated and experimental transition frequencies for radio astronomical observations 
and other purposes are provided 
in the catalog section\footnote{website: https://cdms.astro.uni-koeln.de/classic/entries/, 
see also https://cdms.astro.uni-koeln.de/classic/} of the CDMS.

Since \citet{H2CS_isos_1982} determined the D$_2$CS dipole moment to be slightly larger 
than that of H$_2$CS, we discuss changes of dipole moments upon isotopic substitution. 
The experimentally determined difference between H$_2$CS and D$_2$CS is only 
$0.0105 \pm 0.0011$~D, equivalent to an overestimation of the D$_2$CS column density 
by about 1.3\,\%, which is negligible by astronomical standards. 
\citet{H2CS_dip_1977} determined that the dipole moments of D$_2$CO is 0.0154~D larger 
than that of H$_2$CO whereas the one of D$_2$CCO is only 0.0024~D larger than the one 
of H$_2$CCO, suggesting that dipole moment differences upon deuteration decrease rapidly 
for increasingly larger molecules. Heavy atom substitution leads to much smaller differences. 
As may be expected, the dipole moment of H$_2^{13}$CO is only 0.0002~D larger than that 
of H$_2$CO \citep{H2CS_dip_1977}. Our ground state dipole moments, which we calculated 
at the B3LYP/QZ and MP2/QZ levels, yielded differences of similar magnitude upon heavy 
atom substitution. The difference in the case of D$_2$CS was 0.0136~D and 0.0132~D after 
scaling the values with the ratio between calculated H$_2$CS ground state dipole moment 
and the experimental value. These isotopic changes agree with the experimental one of 
\citet{H2CS_isos_1982} within three times the uncertainty. Finally, we point out that 
the slight rotation of the inertial axis system in the case of HDCS leads to a minute 
$b$-dipole moment component of $\sim$0.08~D. The strongest $b$-type transitions are 
around three orders of magnitude weaker than the strongest $a$-type transitions at 
similar frequencies. Only the uncertainties of the $K_a = 1 \leftrightarrow 0$ transitions 
($\sim$0.2~MHz) may be small enough to permit detection in astronomical spectra at least 
in theory. The uncertainties increase rapidly with increasing $K_a$.

Astronomers will be interested to know the impact of the present data on the calculated 
line positions. Deviations between initial calculations of transition frequencies and 
the present ones increase usually strongly with $K_a$ and less strongly with $J$. 
We show in Table~\ref{deviations} for three isotopic species and selected values 
of $K_a$ the $J$ values and the corresponding frequencies for which these deviations 
exceed 1~MHz. This corresponds to the line width of the protostar IRAS 16293$-$2422 
source B around 300~GHz; e.g., \citet{sulfur_IRAS16293_2018}. 
Dark clouds may exhibit even smaller line widths, whereas high-mass star-forming 
regions usually display larger line widths by factors of a few.

Noticing that initially calculated and present transition frequencies of H$_2$C$^{34}$S 
display differences of more than 1~MHz in the upper millimeter and lower submillimeter region 
for modest values of $K_a$, we wondered if the findings of \citet{sulfur_IRAS16293_2018} were 
affected by these differences. The paper is based on the Protostellar Interferometric Line 
Survey (PILS) of the binary IRAS 16293$-$2422 carried out with the Atacama Large 
Millimeter/submillimeter Array (ALMA) in the 329.15$-$362.90~GHz range \citep{PILS_2016}. 
The $J = 10 - 9$ transitions of H$_2$C$^{34}$S are covered in that survey. The model by 
\citet{sulfur_IRAS16293_2018} shows that the $K_a = 3$ pair of transitions are near 
the noise limit, but are blended by stronger transitions. Shifts of almost 1~MHz between 
initially and presently calculated rest frequencies do not change this enough 
(Drosdovskaya, private communication to H.S.P.M., 2018). 
All but one of the remaining five transitions with lower $K_a$ are clearly blended. 
Therefore, the finding by \citet{sulfur_IRAS16293_2018} that only an upper limit to 
the column density could be determined for H$_2$C$^{34}$S remains unaltered.

We were able to determine $A - (B + C)/2$ for all isotopic species and for some even $D_K$ 
although direct information on the purely $K$-dependent parameters exists only for the main 
species through the $\Delta K_a = 2$ rotational transitions and the ground state combination 
differences from \citet{H2CS_A-X_1994} and even though thioformaldehyde is so close to 
the symmetric prolate limit.

The experimental $^{33}$S hyperfine structure parameters in Table~\ref{tab-HFS-parameters} 
agree well or quite well with those from quantum-chemical calculations. We note that the 
calculated values are equilibrium values whereas the experimental ones are values referring 
to the ground vibrational state. Consideration of vibrational effects may have some influence 
on the comparison, but their evaluation was beyond the aim of our study.

A comparison of spectroscopic parameters of the isovalent molecules formaldehyde, 
thioformaldehyde, silanone, and thiosilanone is given in Table~\ref{tab_comp-parameters}. 
Interestingly, the quartic distortion parameters scale approximately with appropriate 
powers of $A - (B + C)/2$, $B + C$, and $B - C$. This appears to apply also for some 
of the available sextic distortion parameters, but in many cases the relations are more 
complex.


\begin{table*}
\begin{center}
\caption{Ground state rotational parameters $B_{g,0}$ of thioformaldehyde isotopic species, vibrational 
         $\Delta B_{i,{\rm v}}$$^a$, electronic $\Delta B_{i,{\rm el}}$ and centrifugal corrections 
         $\Delta B_{i,{\rm cent}}$, coupled-cluster corrected semi-empirical equilibrium rotational 
         parameters $B_{i,e}$(CCSD(T), and resulting equilibrium inertia defect $\Delta _e$.$^b$}
\label{tab_equi-parameters}
\renewcommand{\arraystretch}{1.10}
\begin{tabular}[t]{llr@{}lr@{}lr@{}lr@{}lr@{}lr@{}lr@{}lr@{}l}
\hline \hline
Species   & $B_i$ & \multicolumn{2}{c}{$B_{i,0}$} & \multicolumn{2}{c}{$\Delta B_{i,{\rm v}}$(B3LYP)} & 
\multicolumn{2}{c}{$\Delta B_{i,{\rm v}}$(MP2)} & \multicolumn{2}{c}{$\Delta B_{i,{\rm v}}$(CCSD(T))} & 
\multicolumn{2}{c}{$\Delta B_{i,{\rm el}}$} & \multicolumn{2}{c}{$\Delta B_{i,{\rm cent}}$} & 
\multicolumn{2}{c}{$B_{i,e}$(CCSD(T)} & \multicolumn{2}{c}{$\Delta _e$} \\
\hline
H$_2$CS             & $A$ & 291613&.34  & 1671&.810 & 1900&.593 & 1958&.684 & 843&.435 & $-$0&.608 & 294414&.850 &     &        \\
                    & $B$ &  17698&.994 &   63&.975 &   68&.066 &   74&.002 &   1&.294 & $-$0&.608 &  17773&.682 &     &        \\
                    & $C$ &  16652&.498 &   98&.197 &  102&.170 &  108&.190 &   0&.218 &    0&.912 &  16761&.817 & $-$0&.000064 \\
H$_2$C$^{33}$S      & $A$ & 291612&.9   & 1672&.606 & 1901&.179 & 1959&.382 & 843&.435 & $-$0&.598 & 294415&.07  &     &        \\
                    & $B$ &  17538&.843 &   63&.279 &   67&.354 &   73&.212 &   1&.282 & $-$0&.598 &  17612&.739 &     &        \\
                    & $C$ &  16510&.639 &   96&.945 &  100&.900 &  106&.845 &   0&.216 &    0&.897 &  16618&.596 & $-$0&.000049 \\
H$_2$C$^{34}$S      & $A$ & 291612&.0   & 1673&.333 & 1901&.704 & 1960&.038 & 843&.435 & $-$0&.588 & 294414&.91  &     &        \\
                    & $B$ &  17388&.498 &   62&.627 &   66&.686 &   72&.472 &   1&.271 & $-$0&.588 &  17461&.653 &     &        \\
                    & $C$ &  16377&.325 &   95&.774 &   99&.712 &  105&.586 &   0&.215 &    0&.882 &  16484&.009 & $-$0&.000030 \\
H$_2$C$^{36}$S      & $A$ & 291609&.7   & 1674&.700 & 1902&.681 & 1961&.244 & 843&.435 & $-$0&.571 & 294413&.75  &     &        \\
                    & $B$ &  17111&.828 &   61&.430 &   65&.458 &   71&.122 &   1&.251 & $-$0&.571 &  17183&.630 &     &        \\
                    & $C$ &  16131&.646 &   93&.630 &   97&.537 &  103&.282 &   0&.211 &    0&.857 &  16235&.996 &    0&.000019 \\
H$_2^{13}$CS        & $A$ & 291628&.8   & 1657&.739 & 1884&.781 & 1943&.937 & 843&.435 & $-$0&.567 & 294415&.62  &     &        \\
                    & $B$ &  16998&.786 &   59&.621 &   63&.806 &   69&.345 &   1&.243 & $-$0&.567 &  17068&.807 &     &        \\
                    & $C$ &  16031&.192 &   91&.529 &   95&.565 &  101&.210 &   0&.210 &    0&.850 &  16133&.462 &    0&.000008 \\
H$_2^{13}$C$^{34}$S & $A$ & 291638&.    & 1658&.041 & 1885&.861 & 1944&.755 & 843&.435 & $-$0&.548 & 294426&.    &     &        \\
                    & $B$ &  16685&.867 &   57&.814 &   61&.982 &   67&.761 &   1&.220 & $-$0&.548 &  16754&.300 &     &        \\
                    & $C$ &  15752&.571 &   89&.600 &   93&.602 &   98&.981 &   0&.206 &    0&.822 &  15852&.580 & $-$0&.000705 \\
HDCS                & $A$ & 202715&.3   &  930&.335 & 1041&.110 & 1070&.464 & 585&.478 & $-$0&.466 & 204370&.79  &     &        \\
                    & $B$ &  16110&.811 &   56&.576 &   58&.665 &   64&.156 &   1&.178 & $-$0&.466 &  16175&.679 &     &        \\
                    & $C$ &  14891&.548 &   88&.063 &   90&.525 &   95&.721 &   0&.195 &    0&.699 &  14988&.163 &    0&.00255  \\
D$_2$CS             & $A$ & 146399&.0   &  645&.743 &  736&.128 &  751&.566 & 422&.765 & $-$0&.369 & 147572&.97  &     &        \\
                    & $B$ &  14904&.267 &   50&.835 &   51&.155 &   56&.344 &   1&.089 & $-$0&.369 &  14961&.330 &     &        \\
                    & $C$ &  13495&.845 &   80&.217 &   80&.936 &   85&.563 &   0&.177 &    0&.554 &  13582&.139 &    0&.00547  \\
\hline
\end{tabular}
\end{center}
\tablefoot{
$^{(a)}$ $\Delta B_{i,{\rm v}} = \sum_{j} \alpha _j^{B_i}$ calculated by different quantum-chemical means 
         as detailed in Sect.~\ref{qcc}. 
$^{(b)}$ All numbers in units of MHz, except $\Delta _e$ in units of amu\,{\AA}$^2$. 
}
\end{table*}

\section{Structural parameters of thioformaldehyde}
\label{structure}

The equilibrium structure is the best and easiest defined structure of a molecule. 
It requires to calculate equilibrium rotational parameter(s), for example, $B_e$ 
from the ground state rotational parameter(s) $B_0$ as follows

\begin{equation}
\label{equi-B}
B_e = B_0 + \frac{1}{2}\sum_{j} \alpha _j^B - \frac{1}{4}\sum_{j \le k} \gamma _{jk}^B - ...
\end{equation}

\noindent
where the $\alpha _j^B$ are first order vibrational corrections, the $\gamma _{jk}^B$ are 
second order vibrational corrections, and so on. Equivalent formulations hold for $A_e$ and 
$C_e$. In the case of a diatomic molecule, only information on one isotopic species is necessary, 
and only one rotational parameter and one vibrational correction of each order exist. Experimental 
data exist for a plethora of diatomic molecules, see, e.g., \citet{const-diat_1979}. $B_e$ is 
usually much larger than $\alpha$ which in turn is much larger than $|\gamma|$; the situation 
involving higher order corrections may be more complex.

The general $n$-atomic asymmetric rotor molecule has three different rotational parameters $A$, 
$B$, and $C$, $3n - 6$ first order vibrational corrections, $(3n - 6)(3n - 5)/2$ second order 
vibrational corrections, and so on. Specifically, the number of first and second order corrections 
are three and six, respectively, for a triatomic molecule, and six and 21 for a tetratomic molecule 
such as H$_2$CS. Experimental equilibrium structural parameters of polyatomic molecules with 
consideration of more than first order vibrational corrections are very rare, but more exist 
with consideration of first order vibrational corrections only. It is necessary to point out 
that $B_0 - B_j$ is only to first order equal to $\alpha _j^B$. Moreover, data for more than 
one isotopic species are needed to determine all independent structural parameters unless 
the molecule is a symmetric triatomic of the type AB$_2$, where atoms A and B do not need 
to be different.

An alternative, lately very common, approach is to calculate $\sum_{j} \alpha _j^B$ by 
quantum-chemical means to derive semi-empirical equilibrium rotational parameters $B_{i,e}$ 
from the experimental ground state values \citep{r_e_emp_1998}. Second and higher order 
vibrational contributions are neglected. Numerous quantum-chemical programs are available 
to carry out such calculations; examples have been mentioned in Sect.~\ref{qcc}.

We have used B3LYP, MP2, and CCSD(T) calculations with an adequately large  basis set of 
quadruple zeta quality to evaluate the first order vibrational corrections for isotopologs 
of thioformaldehyde which have been summarized in Table~\ref{tab_equi-parameters} together 
with ground state values, two additional corrections described in greater detail below, 
the final semi-empirical equilibrium values at the CCSD(T) level, and the corresponding 
equilibrium inertia defect $\Delta _e$. The inertia defect is defined as $\Delta = 
I_{cc} - I_{bb} - I_{aa}$. Among the three methods employed in the present study, CCSD(T) 
is considered to be by far the most accurate one under most circumstances whereas MP2 and 
B3LYP are usually less accurate by different degrees. \citet{H2CO_ai_2018} performed 
extensive calculations on the related formaldehyde molecule. The ae-CCSD(T) data obtained 
with a basis set of QZ quality are already quite close to experimental values, but larger 
basis sets or higher degrees of electron correlation modify the picture somewhat. 
All corrections together improve the agreement between quantum-chemical calculations 
and experimental results non-negligibly. Such calculations are, however, very demanding, 
and in many cases ae-CCSD(T) calculations with a basis set of QZ quality are a good 
compromise for small to moderately large molecules \citep{2nd-row_ai_2005}.

The $\Delta B_{i,{\rm v}}$ determined for thioformaldehyde differ considerably among 
the three methods for each isotopolog and each $i$. We suspect that the second order 
vibrational corrections are smaller than the differences between the methods. 
If we assume that they decrease in magnitude in a similar way as the first order 
corrections are smaller than the ground state rotational parameters, then the second 
order corrections to $A$ should be around 10~MHz, and those to $B$ and $C$ should be 
less than 1~MHz.

\citet{inertia-defects_1961} showed that the rotational Hamiltonian of a semirigid rotor 
contains two terms which cause the inertia defect $\Delta$ to be non-zero when the ground 
state rotational parameters $B_{i,0}$ were corrected for the vibrational corrections 
$\Delta B_{i,{\rm v}}$, namely an electronic contribution $\Delta B_{i,{\rm el}}$ and a 
centrifugal distortion contribution $\Delta B_{i,{\rm cent}}$. The electronic contribution 
is calculated as $\Delta B_{i,{\rm el}} = -B_{i,e}\,g_{ii}\,m_e/m_p$, where the $g_{ii}$ 
are components of the rotational $g$-tensor and $m_e$ and $m_p$ are the masses of the 
electron and the proton, respectively \citep{inertia-defects_1961}. We took the $g_{ii}$ 
values of thioformaldehyde from the very accurate Zeeman measurements of 
\citet{H2CS_Zeeman_1972}. The centrifugal distortion contribution is evaluated as 
$\Delta A_{\rm cent} = \Delta B_{\rm cent} = \hbar ^4\,\tau_{abab}/2$ and 
$\Delta C_{\rm cent} = -3\hbar ^4\,\tau_{abab}/4$ \citep{inertia-defects_1961}. 
$\hbar ^4\,\tau_{abab}$ ($= \tau_{abab}'$; tbc) is a distortion parameter which was evaluated 
here from an empirical force field calculated using the program NCA \citep{nca_1978}.

The inertia defect $\Delta$ may be used an indication of the quality of the vibrational 
correction. The ground state value $\Delta _0$ of the main isotopolog is 
0.06139~amu\,{\AA}$^2$, quite small and positive as can be expected for a small and 
rigid molecule. The equilibrium value should ideally be zero. However, the first order 
vibrational corrections lead usually to negative values which are much smaller in 
magnitude than the ground state values. In the case of our B3LYP, MP2, and CCSD(T) 
calculations, the values are $-$0.00381, $-$0.00305, and $-$0.00406~amu\,{\AA}$^2$, 
respectively. Taking the electronic corrections into account, we obtain 0.00281, 
0.00356, and 0.00255~amu\,{\AA}$^2$, respectively, and finally, after applying 
the centrifugal distortion correction, 0.00020, 0.00093, and 
$-$0.000064~amu\,{\AA}$^2$, respectively.


\begin{table*}
\begin{center}
\caption{Quantum-chemical and experimental bond lengths (pm) and bond angle (deg) of thioformaldehyde.$^a$}
\label{struct-parameters}
\renewcommand{\arraystretch}{1.10}
\begin{tabular}[t]{lr@{}lr@{}lr@{}l}
\hline \hline
Method$^b$ & \multicolumn{2}{c}{$r$(CS)} & \multicolumn{2}{c}{r(CH)} & \multicolumn{2}{c}{$\angle$(HCS)}  \\
\hline
B3LYP/TZ                & 160&.614      & 108&.786      & 122&.202      \\
B3LYP/QZ                & 160&.516      & 108&.711      & 122&.208      \\
B3LYP/awQZ              & 160&.506      & 108&.716      & 122&.198      \\
MP2/TZ                  & 160&.991      & 108&.627      & 121&.881      \\
MP2/QZ                  & 160&.670      & 108&.533      & 121&.825      \\
MP2/awCQZ               & 160&.622      & 108&.550      & 121&.794      \\
ae-MP2/awCQZ            & 160&.188      & 108&.390      & 121&.767      \\
CCSD(T)/TZ              & 161&.826      & 108&.766      & 121&.928      \\
CCSD(T)/QZ              & 161&.415      & 108&.683      & 121&.841      \\
ae-CCSD(T)/wCQZ         & 160&.890      & 108&.531      & 121&.855      \\
ae-CCSD(T)/wC5Z         & 160&.797      & 108&.512      & 121&.815      \\
CCSD(T)/QZ*$^c$         & 160&.90       & 108&.53       & 121&.77       \\
dito, refined$^d$       & 160&.895      & 108&.685      & 121&.75       \\
$r_s$$^e$               & 161&.08~(9)   & 109&.25~(9)   & 121&.57~(3)   \\
$r_s$$^f$               & 161&.077~(1)  & 108&.692~(3)  & 121&.74~(2)   \\
$r_z$$^f$               & 161&.38~(4)   & 109&.62~(6)   & 121&.87~(5)   \\
$r_z$$^g$               & 161&.57~(8)   & 109&.92~(21)  & 121&.33~(29)  \\
$r_e(r_z)$$^g$          & 161&.10~(8)   & 108&.56~(21)  &    &          \\
$r_{I,\epsilon}$        & 161&.025~(30) & 109&.246~(21) & 121&.562~(12) \\
$r_e^{\rm SE}$(B3LYP)   & 160&.975~(2)  & 108&.526~(6)  & 121&.706~(5)  \\
$r_e^{\rm SE}$(MP2)     & 160&.934~(6)  & 108&.556~(15) & 121&.759~(13) \\
$r_e^{\rm SE}$(CCSD(T)) & 160&.909~(1)  & 108&.531~(2)  & 121&.758~(2)  \\

\hline
\end{tabular}
\end{center}
\tablefoot{
$^a$ All values from this work unless indicated otherwise. Numbers in parentheses are one 
     standard deviation in units of the least significant figures.
$^b$ Quantum-chemical calculations as detailed in Sect.~\ref{qcc}. 
$^c$ CCSD(T) calculation with basis sets up to QZ quality with extrapolation to infinite 
     basis set size and with several corrections \citep{H2CS_ai_2011}. 
$^d$ Calculated rotational energies from \citet{H2CS_ai_2011} were adjusted to experimental 
     energies by refining the structural parameters \citep{H2CS_ai_egy-etc_2013}.
$^e$ Substitution structure $r_s$ for H$_2$CS isotopolog from \citet{H2CS_isos_rot_1971}. 
$^f$ Substitution structure $r_s$ and ground state average structure $r_z$ 
     for H$_2$CS isotopolog from \citet{H2CS_isos_1982}. 
$^g$ Ground state average structure $r_z$ for H$_2$CS isotopolog and estimate of 
     equilibrium bond lengths from $r_z$ \citep{H2CS_D2CS_IR_re-est_1981}.
}
\end{table*}

The equilibrium inertia defects in Table~\ref{tab_equi-parameters} show very small scatter 
very close to zero among five isotopic species, and slightly larger scatter for three others. 
Even though that larger scatter is still fairly small, it is worthwhile to look into 
potential sources for that finding. The smaller list of experimental lines could be 
an explanation for H$_2^{13}$C$^{34}$S and for D$_2$CS, but not likely for HDCS. 
The $\Delta _e$ value of H$_2^{13}$C$^{34}$S would be essentially zero if $A_e$ were 
increased by 121~MHz. This can be ruled out safely because ideally $A_e$ should not change 
upon substitution of one (or both) of the heavy atoms, and the H$_2^{13}$C$^{34}$S value 
is only about 11~MHz larger than that of the main isotopolog albeit with an uncertainty 
of 16~MHz. A decrease of $C_e$ by 0.35~MHz would also lead to $\Delta _e \approx 0$, but 
a corresponding change in the experimentally determined value of $C_0$ appears rather unlikely. 
We suspect that shortcomings in the CCDS(T) first order vibrational correction or the neglect 
of second order vibrational correction are mainly responsible for the somewhat larger scatter 
observed for three of the thioformaldehyde isotopologs, even more so, as the differences 
between the equilibrium inertia defects of H$_2$CS and H$_2^{13}$C$^{34}$S are about twice 
as large if the CCDS(T) vibrational corrections are replaced by the B3LYP or MP2 corrections.

We employed the RU111J program \citep{structure_rev_1995} to derive semi-empirical equilibrium 
structural parameters $r_e^{\rm SE}$ as well as $r_{I,\epsilon}$ parameters. The latter model 
was proposed by \citet{r_I-eps_1991}. The difference between ground state and equilibrium 
moments of inertia can be expressed as $I_{ii,0} = I_{ii,e} + \epsilon _i$, with $i = a, b, c$, 
assuming that the $\epsilon _i$ are equal among the different isotopologs 
of a given molecule. According to \citet{r_I-eps_1991}, the $r_{I,\epsilon}$ parameters are 
equivalent with $r_{\Delta I}$ parameters (isotopic differences are fit to determine 
structural parameters) and with substitution parameters $r_s$ (isotopic differences 
between one reference isotopolog and one isotopolog in which one atom has been substituted 
are used to locate that atom). The advantage of considering the $\epsilon _i$ explicitely in 
the calculations are predictions of rotational parameters of isotopic species to be studied, 
see for example \citet{use_r_I-eps_c-PrGeH3_1992} and \citet{use_r_I-eps_SO2ClF_1994}. 
The resulting structural parameters are given in 
Table~\ref{struct-parameters} together with earlier $r_s$ parameters, ground state average 
($r_z$) parameters and an approximation of the equilibrium structure derived from $r_z$ 
parameters. The harmonic contributions to the ground state moments of inertia, obtained from 
a harmonic force field calculation, are subtracted off in the ground state average structure. 
The approximation of the equilibrium structure derived from $r_z$ parameters assumes that 
anharmonic contributions to a given bond in a molecule can be approximated from the anharmonicity 
of the respective diatomic molecule, and differences in $r_z$ and $r_e$ bond angles are 
neglected \citep{H2CS_D2CS_IR_re-est_1981}. 
Table~\ref{struct-parameters} also contains structural parameters of thioformaldehyde 
from several present and selected earlier quantum-chemical calculations.

The semi-empirical strutures $r_e^{\rm SE}$ determined with first order vibrational corrections 
obtained with three different methods are quite similar, albeit with some of the differences 
outside the combined uncertainties. The semi-empirical structure obtained with the CCSD(T) 
corrections is very close to the purely quantum-chemically derived ae-CCSD(T)/wQZ structure, 
as is very often the case \citep{2nd-row_ai_2005}, and is probably closest to a purely 
experimental equilibrium structure. The CS bond lengths derived from B3LYP or MP2 calculations 
with basis sets of QZ quality are all too short, especially the ae-MP2 value. 
The CH bond lengths are all slightly too long, and the HCS bond angles all too large. 
The corresponding MP2 quantities are closer to our semi-empirical values. 

Our $r_{I,\epsilon}$ parameters agree within combined uncertainties with the $r_s$ parameters 
of \citet{H2CS_isos_rot_1971}, as is expected \citep{r_I-eps_1991}, but less so with the 
$r_s$ values of \citet{H2CS_isos_1982}. However, these latter $r_s$ values are quite close 
to our $r_e^{\rm SE}$ values. The two sets of ground state average ($r_z$) parameters 
differ somewhat, but in both cases both bond lengths are longer than the equilibrium values, 
as is usually the case. The equilibrium bond lengths derived from one of the $r_z$ structures 
is in fairly good agreement with our $r_e^{\rm SE}$ values.

We recommend employment of first-order vibrational corrections obtained with the CCSD(T) method 
for semi-empirical structure determinations if high accuracy is desired. Less expensive methods 
may, however, be sufficient if accuracy requirements are less stringent.

\section{Conclusion and outlook}
\label{conclusion}

We have obtained extensive sets of accurate transition frequencies for seven isotopic species 
of thioformaldehyde. They extend to beyond 900~GHz for H$_2$C$^{33}$S, for H$_2$C$^{36}$S, 
and for HDCS and even reach almost 1400~GHz in the cases of H$_2$CS, H$_2$C$^{34}$S, and 
H$_2^{13}$CS. The line list of the very rare H$_2^{13}$C$^{34}$S extends to about 360~GHz. 
The resulting accurate spectroscopic parameters not only permit prediction of the strong 
$R$-branch transitions in the respective frequency range and up to $K_a$ slightly beyond 
those covered in the line lists, but also permit reliable to reasonable extrapolation up 
to about twice the upper experimental frequencies and probably up to $K_a$ covered 
in the line lists. 
Thus, accurate rest frequencies covering the entire present frequency range of ALMA are 
available for most thioformaldehyde isotopologs; in the case of H$_2^{13}$C$^{34}$S, they 
cover all bands up to band~9. In addition, the $^{33}$S hyperfine structure of H$_2$C$^{33}$S 
has been reevaluated based on previous and present data.

We carried out quantum-chemical calculations to evaluate first order vibrational corrections 
to the ground state rotational parameters in order to approximate equilibrium rotational 
parameters which lead to semi-empirical structural parameters. Quantum-chemical calculations 
were also carried out to obtain structural parameters directly.

Additional observed rest frequencies include, for example, data for excited vibrational 
states of H$_2$CS and H$_2$C$^{34}$S. We intend to report on these findings in a separate 
manuscript elsewhere in the near future.


\begin{acknowledgements}
We acknowledge support by the Deutsche Forschungsgemeinschaft via the collaborative 
research centers SFB~494 (project E2) and SFB~956 (project B3) as well as 
the Ger{\"a}tezentrum SCHL~341/15-1 (``Cologne Center for Terahertz Spectroscopy''). 
We are grateful to NASA for its support of the OSU program in laboratory astrophysics 
and the ARO for its support of the study of large molecules. HSPM thanks C.~P. Endres 
and M. Koerber for support during some of the measurements in K{\"o}ln. Our research 
benefited from NASA's Astrophysics Data System (ADS).
\end{acknowledgements}


\section*{Appendix A. Supplementary material}
\label{Appendix}


\begin{table*}
\begin{center}
\caption{Assigned transitions for the H$_2$C$^{34}$S isotopic species as an example, 
observed transition frequencies (MHz)$^a$, experimental uncertainties Unc. (MHz)$^a$, 
residual O$-$C between observed frequencies and those calculated from the final 
set of spectroscopic parameters (MHz)$^a$, weight for blended lines, 
and sources of lines.}
\label{supplement}
\begin{tabular}{rrrrrrrrr@{}llrrl}
\hline \hline
$J'$ & $K_a'$ & $K_c'$ & $F' + 0.5$ & $J''$ & $K_a''$ & $K_c''$ & $F'' + 0.5$ & 
\multicolumn{2}{c}{Frequency} & Unc. & O$-$C & Weight & Source\\
\hline
  6 &  1 &  5 &   &  6 & 1 &  6 &    &   21230&.15   & 0.05  & $-$0.01256 &        & \citet{H2CS_isos_1982}     \\
  7 &  1 &  6 &   &  7 & 1 &  7 &    &   28304&.63   & 0.05  &    0.01393 &        & \citet{H2CS_isos_1982}     \\
  1 &  0 &  1 &   &  0 & 0 &  0 &    &   33765&.80   & 0.05  &    0.05051 &        & \citet{H2CS_isos_1982}     \\
  8 &  1 &  7 &   &  8 & 1 &  8 &    &   36388&.01   & 0.05  & $-$0.07933 &        & \citet{H2CS_isos_1982}     \\
  2 &  1 &  2 &   &  1 & 1 &  1 &    &   66517&.88   & 0.10  & $-$0.02248 &        & \citet{H2CS_isos_rot_1971} \\
  2 &  0 &  2 &   &  1 & 0 &  1 &    &   67528&.15   & 0.10  & $-$0.11515 &        & \citet{H2CS_isos_rot_1971} \\
  2 &  1 &  1 &   &  1 & 1 &  0 &    &   68539&.94   & 0.16  & $-$0.23319 &        & \citet{H2CS_isos_rot_1971} \\
 15 &  1 & 14 &   & 15 & 1 & 15 &    &  121120&.1500 & 0.100 & $-$0.03361 &        & OSU                        \\
  4 &  1 &  4 &   &  3 & 1 &  3 &    &  133026&.9097 & 0.050 & $-$0.01021 &        & OSU                        \\
  4 &  3 &  2 &   &  3 & 3 &  1 &    &  135027&.8171 & 0.050 &    0.01347 & 0.5000 & OSU                        \\
  4 &  3 &  1 &   &  3 & 3 &  0 &    &  135027&.8171 & 0.050 &    0.01347 & 0.5000 & OSU                        \\
  4 &  0 &  4 &   &  3 & 0 &  3 &    &  135030&.6546 & 0.050 & $-$0.00998 &        & OSU                        \\
    &    &    &   &    &   &    &    &        &      &       &            &        &                            \\
 41 &  6 & 36 &   & 40 & 6 & 35 &    & 1378757&.4655 & 0.010 & $-$0.00721 & 0.5000 & Koeln                      \\
 41 &  6 & 35 &   & 40 & 6 & 34 &    & 1378757&.4655 & 0.010 & $-$0.00721 & 0.5000 & Koeln                      \\
 41 &  4 & 38 &   & 40 & 4 & 37 &    & 1380675&.2953 & 0.010 &    0.00481 &        & Koeln                      \\
 41 &  4 & 37 &   & 40 & 4 & 36 &    & 1380920&.5593 & 0.010 &    0.00223 &        & Koeln                      \\
 41 &  3 & 39 &   & 40 & 3 & 38 &    & 1380944&.1970 & 0.010 & $-$0.00340 &        & Koeln                      \\
 42 &  1 & 42 &   & 41 & 1 & 41 &    & 1385516&.6458 & 0.010 & $-$0.02038 &        & Koeln                      \\
\hline
\end{tabular}
\end{center}
\tablefoot{
This table as well as those of other isotopologs are available in their 
entirety in the electronic edition in the online journal: 
http://cdsarc.ustrasbg.fr/cgi-bin/VizieR?-source=J/A+A/Vol/Num. 
A portion is shown here for guidance regarding its form and content. 
The $F$ quantum numbers are redundant for all species except for H$_2$C$^{33}$S. 
$^{(a)}$ Negative uncertainties in the line list of the main isotopic species signal 
that units are cm$^{-1}$ instead of MHz.
}
\end{table*}



\begin{thebibliography}{}

\bibitem[Ag{\'u}ndez et al.(2008)]{H2CS_etc_CW-Leo_2008} 
Ag{\'u}ndez, M., Fonfr{\'{\i}}a, J.~P., Cernicharo, J., Pardo, J.~R., \& Gu{\'e}lin, M. 
2008, \aap, 479, 493 

\bibitem[Bailleux et al.(1994)]{H2SiO_rot_1994} 
Bailleux, S., Bogey, M., Demuynck, C., Destombes, J.-L., \& Walters, A. 
1994, \jcp, 101, 2729 

\bibitem[Becke(1993)]{Becke_1993}
Becke, A.~D. 
1993, \jcp, 98, 5648  

\bibitem[Beers et al.(1972)]{H2CS_rot_1972} 
Beers, Y., Klein, G.~P., Kirchhoff, W.~H., \& Johnson, D.~R. 
1972, J. Mol. Spectrosc., 44, 553 

\bibitem[Belov et al.(1995)]{CO-BWO_1995} 
Belov, S.~P., Lewen, F., Klaus, T., \& Winnewisser, G. 
1995, J. Mol. Spectrosc., 174, 606 

\bibitem[Berglund \& Wieser(2011)]{iso-comp_2011} 
Berglund, M., \& Wieser, M. E. 
2011, Pure Appl. Chem., 83, 397

\bibitem[Brown et al.(1987)]{H2CS_HFS_1987} 
Brown, R.~D., Godfrey, P.~D., McNaughton, D., \& Yamanouchi, K. 
1987, Mol. Phys., 62, 1429 

\bibitem[Christen(1978)]{nca_1978}
Christen, D. 
1978, J. Mol. Struct., 48, 101

\bibitem[Clouthier et al.(1994)]{H2CS_A-X_1994} 
Clouthier, D.~J., Huang, G., Adam, A.~G., \& Merer, A.~J. 
1994, \jcp, 101, 7300 

\bibitem[Coriani et al.(2005)]{2nd-row_ai_2005} 
Coriani, S., Marchesan, D., Gauss, J., et al. 
2005, \jcp, 123, 184107 

\bibitem[Cox et al.(1982)]{H2CS_isos_1982} 
Cox, A.~P., Hubbard, S.~D., \& Kato, H. 
1982, J. Mol. Spectrosc., 93, 196

\bibitem[Crockett et al.(2014)]{Orion-KL_HIFI_2014} 
Crockett, N.~R., Bergin, E.~A., Neill, J.~L., et al. 
2014, \apj, 787, 112 

\bibitem[Cummins et al.(1986)]{SgrB2_survey_1986}  
Cummins, S.~E., Linke, R.~A., \& Thaddeus, P. 
1986, \apjs, 60, 819 

\bibitem[De Lucia(2010)]{sub-mmW_2010} 
De Lucia, F.~C. 
2010, J. Mol. Spectrosc., 261, 1 

\bibitem[Drozdovskaya et al.(2018)]{sulfur_IRAS16293_2018} 
Drozdovskaya, M.~N., van Dishoeck, E.~F., J{\o}rgensen, J.~K., et al. 
2018, \mnras, 476, 4949 

\bibitem[Dubernet et al.(2010)]{VAMDC_2010}
Dubernet, M.~L., Boudon, V., Culhane, J.~L., et al. 
2010, J. Quant. Spectrosc. Radiat. Transfer, 111, 2151

\bibitem[Dubernet et al.(2016)]{VAMDC_2016} 
Dubernet, M.~L., Antony, B.~K., Ba, Y.~A., et al. 
2016, J. Phys. B, 49, 074003 

\bibitem[Dunning(1989)]{cc-pVXZ_1989}   
Dunning, T.~H., Jr.
1989, \jcp, 90, 1007 

\bibitem[Dunning et al.(2001)]{plusD_1995}
Dunning, T.~H., Jr., Peterson, K.~A., \& Wilson, A.~K. 
2001, \jcp, 114, 9244  

\bibitem[Endres et al.(2016)]{CDMS_3}
Endres, C.~P., Schlemmer, S., Schilke, P., Stutzki, J., \& M{\"u}ller, H.~S.~P. 
2016, J. Mol. Spectrosc., 327, 95 

\bibitem[Epple \& Rudolph(1992)]{use_r_I-eps_c-PrGeH3_1992} 
Epple, K.~J. \& Rudolph, H.~D. 
1992, J. Mol. Spectrosc., 152, 355

\bibitem[Fabricant et al.(1977)]{H2CS_dip_1977} 
Fabricant, B., Krieger, D., \& Muenter, J.~S. 
1977, \jcp, 67, 1576 

\bibitem[Frisch et al.(2013)]{G09E}
Gaussian 09, Revision E.01,
Frisch,  M.~J., Trucks, G.~W., Schlegel, H.~B., et al., 
Gaussian, Inc., Wallingford CT, 2013. 

\bibitem[Gardner et al.(1985)]{H2CS-34_1985}  
Gardner, F.~F., Hoglund, B., Shukre, C., Stark, A.~A., \& Wilson, T.~L. 
1985, \aap, 146, 303 

\bibitem[Groner et al.(2007)]{dispersion-corr_2007} 
Groner, P., Winnewisser, M., Medvedev, I.~R., et al. 
2007, \apjs, 169, 28 

\bibitem[Heikkil{\"a} et al.(1999)]{LMC_SMC_1999}  
Heikkil{\"a}, A., Johansson, L.~E.~B., \& Olofsson, H. 
1999, \aap, 344, 817 

\bibitem[Huber \& Herzberg(1979)]{const-diat_1979} 
Huber, K.~P. \& Herzberg, G. 
Molecular Spectra and Molecular Structure: IV. Constants of Diatomic Molecules. 
1979, Van Nostrand Reinhold, New York

\bibitem[Irvine et al.(1989)]{H2CS_dark-clouds_1989}  
Irvine, W.~M., Friberg, P., Kaifu, N., et al. 
1989, \apj, 342, 871 

\bibitem[Johnson \& Powell(1970)]{H2CS_rot_1970} 
Johnson, D.~R., \& Powell, F.~X. 
1970, Science, 169, 679 

\bibitem[Johnson et al.(1971)]{H2CS_isos_rot_1971} 
Johnson, D.~R., Powell, F.~X., \& Kirchhoff, W.~H. 
1971, J. Mol. Spectrosc., 39, 136 

\bibitem[J{\o}rgensen et al.(2016)]{PILS_2016} 
J{\o}rgensen, J.~K., van der Wiel, M.~H.~D., Coutens, A., et al. 
2016, \aap, 595, A117 

\bibitem[Lee et al.(1988)]{LYP_1988}
Lee, C., Yang, W., \& Parr, R.~G. 
1988, Phys. Rev. B, 37, 785  

\bibitem[Maeda et al.(2006)]{dispersion-corr_2006} 
Maeda, A., De Lucia, F.~C., Herbst, E., et al. 2006, \apjs, 162, 428 

\bibitem[Maeda et al.(2008)]{H2CS_rot_2008} 
Maeda, A., Medvedev, I.~R., Winnewisser, M., et al. 
2008, \apjs, 176, 543

\bibitem[Marcelino et al.(2005)]{D2CS_det_2005} 
Marcelino, N., Cernicharo, J., Roueff, E., Gerin, M., \& Mauersberger, R. 
2005, \apj, 620, 308 

\bibitem[Martin et al.(1994)]{H2CS_FF_ai_1994} 
Martin, J.~M.~L., Francois, J.~P., \& Gijbels, R. 
1994, J. Mol. Spectrosc., 168, 363 

\bibitem[Mart{\'{\i}}n et al.(2005)]{NGC253_2005} 
Mart{\'{\i}}n, S., Mart{\'{\i}}n-Pintado, J., Mauersberger, R., Henkel, C., \& Garc{\'{\i}}a-Burillo, S. 
2005, \apj, 620, 210 

\bibitem[McCarthy et al.(2011)]{H2SiS_Si2S_rot_2011} 
McCarthy, M.~C., Gottlieb, C.~A., Thaddeus, P., Thorwirth, S., \& Gauss, J. 
2011, \jcp, 134, 034306 

\bibitem[McNaughton \& Bruget(1993)]{H2CS_FIR_nu2_1993} 
McNaughton, D., \& Bruget, D.~N. 
1993, J. Mol. Spectrosc., 159, 340 

\bibitem[Medvedev et al.(2004)]{DEE_rot_FASSST_2004} 
Medvedev, I., Winnewisser, M., De Lucia, F.~C., et al. 
2004, J. Mol. Spectrosc., 228, 314

\bibitem[Medvedev et al.(2005)]{CAAARS_2005} 
Medvedev, I.~R., Winnewisser, M., Winnewisser, B.~P., De Lucia, F.~C., \& Herbst, E. 
2005, J. Mol. Struct., 742, 229 

\bibitem[Mills(1972)]{vib-rot_rev_1972} 
Mills, I.~M., ``Vibration-Rotation Structure in Asymmetric- and Symmetric-Top Molecules'', 
in ``Modern Spectroscopy: Modern Research'', Rao, K.~N. \& Matthews, C.~W., 
eds., Academic Press, New York, NY, USA, vol.~I, 1972; pp. 115$-$140

\bibitem[Minowa et al.(1997)]{HDCS_rot_det_1997} 
Minowa, H., Satake, M., Hirota, T., et al. 
1997, \apjl, 491, L63 

\bibitem[M{\o}ller \& Plesset(1934)]{MPn_1934}
M{\o}ller, C. \& Plesset, M.~S. 
1934, Phys. Rev., 46, 618  

\bibitem[Morgan et al.(2018)]{H2CO_ai_2018} 
Morgan, W.~J., Matthews, D.~A., Ringholm, M., et al. 
2018, J. Chem. Theory Comput., 14, 1333

\bibitem[Muller et al.(2011)]{PKS1830_4mm_2011}  
Muller, S., Beelen, A., Gu{\'e}lin, M., et al. 
2011, \aap, 535, A103 

\bibitem[M{\"u}ller \& Gerry(1994)]{use_r_I-eps_SO2ClF_1994} 
M{\"u}ller, H.~S.~P., \& Gerry, M.~C.~L. 
1994, J. Chem. Soc. Faraday Trans., 90, 2601 

\bibitem[M{\"u}ller \& Br{\"u}nken(2005)]{SO2_rot_2005} 
M{\"u}ller, H.~S.~P., \& Br{\"u}nken, S. 
2005, J. Mol. Spectrosc., 232, 213 

\bibitem[M{\"u}ller \& Lewen(2017)]{H2CO-X_rot_2017} 
M{\"u}ller, H.~S.~P., \& Lewen, F. 
2017, J. Mol. Spectrosc., 331, 28 

\bibitem[M{\"u}ller et al.(2007)]{SiS_rot_2007} 
M{\"u}ller, H.~S.~P., McCarthy, M.~C., Bizzocchi, L., et al. 
2007, Phys. Chem. Chem. Phys., 9, 1579 

\bibitem[M{\"u}ller et al.(2015)]{MeCN_v8_le_2_rot_2015} 
M{\"u}ller, H.~S.~P., Brown, L.~R., Drouin, B.~J., et al. 
2015, J. Mol. Spectrosc., 312, 22 

\bibitem[M{\"u}ller et al.(2016)]{MeCN_13C-vib_rot_2016} 
M{\"u}ller, H.~S.~P., Drouin, B.~J., Pearson, J.~C., et al. 
2016, \aap, 586, A17 

\bibitem[Neill et al.(2014)]{SgrB2N_HIFI_2014} 
Neill, J.~L., Bergin, E.~A., Lis, D.~C., et al. 
2014, \apj, 789, 8 

\bibitem[Oka \& Morino(1961)]{inertia-defects_1961} 
Oka, T., \& Morino, Y. 
1961, J. Mol. Spectrosc., 6, 472 

\bibitem[Olofsson et al.(2007)]{Orion-KL-survey_2007-1} 
Olofsson, A.~O.~H., Persson, C.~M., Koning, N., et al. 
2007, \aap, 476, 791 

\bibitem[Persson et al.(2007)]{Orion-KL-survey_2007-2} 
Persson, C.~M., Olofsson, A.~O.~H., Koning, N., et al. 
2007, \aap, 476, 807 

\bibitem[Peterson \& Dunning(2002)]{core-corr_2002}
Peterson, K.~E. \& Dunning, T.~H., Jr.
2002, \jcp, 117, 10548 

\bibitem[Petkie et al.(1997)]{FASSST_1997} 
Petkie, D.~T., Goyette, T.~M., Bettens, R.~P.~A., et al. 
1997, Rev. Sci. Instrum., 68, 1675 

\bibitem[Pickett(1991)]{spfit_1991} 
Pickett, H.~M. 
1991, J. Mol. Spectrosc., 148, 371

\bibitem[Raghavachari et al.(1989)]{CC+T_1989}
Raghavachari, K., Trucks, G.~W., Pople, J.~A. \& Head-Gordon, M. 
1989, Chem. Phys. Lett., 157, 479

\bibitem[Rock \& Flygare(1972)]{H2CS_Zeeman_1972} 
Rock, S.~L., \& Flygare, W.~H. 
1972, \jcp, 56, 4723 

\bibitem[Rudolph(1991)]{r_I-eps_1991} 
Rudolph, H.~D. 
1991, Struct. Chem, 2, 581

\bibitem[Rudolph(1995)]{structure_rev_1995} 
Rudolph, H.~D., ``Accurate Molecular Structure from Microwave Rotational Spectroscopy'', 
in ``Advances in Molecular Structure Research'', Hargittai, I. \& Hargittai, M., 
eds., JAI Press Inc., Greenwich, CT, USA, vol.~I, 1995

\bibitem[Schilke et al.(1997)]{Orion-KL-survey_1997}  
Schilke, P., Groesbeck, T.~D., Blake, G.~A., Phillips, \& T.~G. 
1997, \apjs, 108, 301 

\bibitem[Schilke et al.(2001)]{Orion-KL-survey_2001}  
Schilke, P., Benford, D.~J., Hunter, T.~R., Lis, D.~C., \& Phillips, T.~G. 
2001, \apjs, 132, 281 

\bibitem[Sinclair et al.(1973)]{H2CS_det_1973} 
Sinclair, M.~W., Fourikis, N., Ribes, J.~C., et al. 
1973, Aust. J. Phys., 26, 85 

\bibitem[Stanton et al.(1998)]{r_e_emp_1998} 
Stanton, J.~F., Lopreore, C.~L., \& Gauss, J. 
1998, \jcp, 108, 7190

\bibitem[Turner et al.(1981)]{H2CS_D2CS_IR_re-est_1981} 
Turner, P.~H., Halonen, L., \& Mills, I.~M. 
1981, J. Mol. Spectrosc., 88, 402 

\bibitem[Winnewisser et al.(1994)]{BWO-review_1994} 
Winnewisser, G., Krupnov, A.~F., Tretyakov, M.~Y., et al. 
1994, J. Mol. Spectrosc., 165, 294 

\bibitem[Winnewisser(1995)]{BWO-review_1995} 
Winnewisser, G. 
1995, Vib. Spectrosc., 8, 241.

\bibitem[Woodney et al.(1997)]{H2CS_Hale-Bopp_1997} 
Woodney, L.~M., A'Hearn, M.~F., McMullin, J., \& Samarasinha, N. 
1997, Earth Moon and Planets, 78, 69

\bibitem[Xu et al.(2012)]{CH3SH_rot_THz-chain_2012} 
Xu, L.-H., Lees, R.~M., Crabbe, G.~T., et al. 
2012, \jcp, 137, 104313

\bibitem[Yachmenev et al.(2011)]{H2CS_ai_2011} 
Yachmenev, A., Yurchenko, S.~N., Ribeyre, T., \& Thiel, W. 
2011, \jcp, 135, 074302 

\bibitem[Yachmenev et al.(2013)]{H2CS_ai_egy-etc_2013} 
Yachmenev, A., Polyak, I., \& Thiel, W. 
2013, \jcp, 139, 204308 



\end{thebibliography}
\end{document}